\documentclass[twocolumn]{aastex63}

\pdfoutput=1

\usepackage{amssymb}
\usepackage{amsmath}
\usepackage{amsfonts}
\usepackage{graphicx}
\usepackage{color}
\usepackage{hyperref}
\usepackage[caption=false]{subfig}
\usepackage{mathtools}

\def\lsim{~\rlap{$<$}{\lower 1.0ex\hbox{$\sim$}}}
\def\bsim{~\rlap{$>$}{\lower 1.0ex\hbox{$\sim$}}}
\def\kms{\ {\rm km\,s^{-1}}}

\def\hkpc{\ {\rm {\it h}^{-1}kpc}}
\def\hmpc{\ {\rm {\it h}^{-1}Mpc}}

\def\hmsun{\ {\rm {\it h}}^{-1}M_\odot}

\def\Kel{\ {\rm K}}

\def\apjl{Astrophys.\ J.\ Lett.}

\def\aap{Astron.\ Astrophys.}

\def\apjs{Astrophys.\ J. Supp.}

\allowdisplaybreaks

\usepackage[normalem]{ulem}

\begin{document}

\title{Low-ionization Metal Absorption at $0.7 \lesssim z \lesssim 2$ \\
Confronting Cosmological Simulations with Observations}

\author[0009-0007-7887-783X]{Ivan Rapoport}
\email{ivanr@campus.technion.ac.il}
\affiliation{Physics Department, Technion -- Israel Institute of Technology, Haifa 3200003, Israel}

\author[0000-0001-9735-4873]{Ehud Behar}
\affiliation{Physics Department, Technion -- Israel Institute of Technology, Haifa 3200003, Israel}

\author[0000-0003-2062-8172]{Vincent Desjacques}
\affiliation{Physics Department, Technion -- Israel Institute of Technology, Haifa 3200003, Israel}

\date{\today}

\begin{abstract}
Low-ionization metal absorption lines provide a primary probe of cool gas in and around galaxies. We confront observations of metal-line absorption in quasar spectra with predictions from the IllustrisTNG cosmological simulation in order to benchmark how well current galaxy formation models reproduce the observed circumgalactic medium (CGM) and intergalactic medium (IGM) absorption signatures. We implement two ionization prescriptions: a purely collisional model 
and a model including photo-ionization by a uniform ultraviolet background (UVB). 
Using a grid-based framework, we compute Mg\,I, Mg\,II and Fe\,II column densities and construct column density probability distribution functions (PDFs) and equivalent width (EW) statistics for comparison with observations. The observational samples considered here are based on the High Resolution Echelle Spectrometer (HIRES), the Ultraviolet and Visual Echelle Spectrograph (UVES), the Sloan Digital Sky Survey (SDSS) and the Dark Energy Spectroscopic Instrument (DESI). 
The computed PDFs broadly reproduce the observed ones 
across the sampled column density range of $10^{11.4}\lesssim \text{N}\lesssim 10^{16}\ \rm{cm^{-2}}$, indicating that the simulation captures the dominant physical drivers of low-ionization
absorption. 
We then compute the cosmic incidence of Mg\,II systems, namely the evolution of their number with redshift $d\mathcal{N}/{dz}$. 
The model that includes UVB accurately produces $d\mathcal{N}/{dz}$ up to equivalent widths (EW) of $\rm W_0^{2796} < 0.6\,\text{\AA}$, consistent with low-density photo-ionized gas in the outer CGM. 
At high EWs of $\rm W_0^{2796} > 1\ \text{\AA}$ TNG underestimates $d\mathcal{N}/{dz}$ and fails to capture its rise toward $z\sim2$. \\
\end{abstract}

\section{Introduction}
\label{sec:intro}
Quasar absorption spectroscopy has long been one of the most powerful tools for studying the diffuse gas and the underlying large scale distribution of baryons in the Universe \citep{Weymann_1979, Young_1982, Bergeron_1991, press/etal:1993, Jannuzi_1998,Rauch_1998, Kim_2013}. As the light from distant quasars passes through intervening structures, it imprints a series of absorption features that record the distribution and chemical composition of the circumgalactic (CGM) and intergalactic (IGM) media \citep{gunn/peterson:1965, bahcall/salpeter:1965, Meiksin_2009_IGM, Tumlinson_2017}. Large spectroscopic surveys—most notably SDSS \citep{York_2000} and its extensions \citep{Eisenstein_2011,Dawson_2013,Lyke_2020} have transformed this technique from targeted studies of individual systems into a statistical science, enabling population-level measurements of absorption systems across cosmic time \citep{Prochaska_2008,Noterdaeme_2012,Pieri_2014,Lan_2017}.

Low-ionization metal absorption lines provide a powerful probe of the cool, enriched gas in and around galaxies. Arising primarily in material with temperatures $T \sim 10^{4-4.5}\Kel$, ions such as Mg\,II, Fe\,II, C\,II and Si\,II trace the cool phase of the CGM as well as the interfaces between galaxies and the IGM \citep{Bergron_Stasinska_1986, Steidel_1992, Churchill_2000,Kacprzak_2008,Weiner_2009,Bouche_2012,Werk_2013,Lan2025}. 
Among these species, Mg\,II is particularly well suited for observational studies owing to its $\lambda\lambda2796,2803$ doublet. 
The distinctive doublet structure combined with relatively strong oscillator strengths makes it readily detectable in ground-based spectra over a broad redshift range \citep{Nestor_2005, Chen_2010}. Absorption by Fe\,II, while more challenging to constrain due to its numerous transitions, provides complementary information on the physical conditions of the gas \citep{Churchill_2000}.

Statistical analyses of Mg\,II absorption systems have revealed a rich population whose properties evolve with cosmic time. Early surveys established the redshift dependence of strong absorbers and characterized their equivalent-width (EW) distribution \citep{Nestor_2005, Zhu_2013, Chen_2017, Churchill2025}, showing that the incidence rate of systems with rest-frame EWs $\gtrsim0.5\ \text{\AA}$ increases toward $z\approx2$ and persists to even higher redshifts \citep{Nestor_2005,Matejek_2012,Seyffert_2013}. Subsequent work extended these measurements to weaker absorbers, demonstrating a continuous distribution spanning more than an order of magnitude in EW and indicating a broad range of physical origins \citep{Rigby_2002, Narayanan_2005, Churchill_2003, Quider_2011, Lan_2017}. Clustering analyses further revealed that typical Mg\,II absorbers occupy dark-matter halos of mass $\langle \log M/M_\odot\rangle \approx 12.1$ \citep{Gauthier_2014}, while their kinematics and spatial correlations point to contributions from outflows and inflows of cool CGM gas \citep{Bouche_2012, Weiner_2009}. Despite these advances, the precision of current statistics remains limited by heterogeneous spectral resolution and signal-to-noise across existing surveys, which particularly affect the detectability of weak systems and introduce redshift-dependent selection biases \citep{Zhu_2013, Lan_2014}. High-resolution spectroscopic samples, such as the HIRES/UVES compilation of \citet{Churchill_2020}, provide an important dataset by resolving absorber kinematics and enabling more accurate column density measurements, particularly for weak systems.

DESI \citep{desicollaboration2016,Chaussidon_2023,DESI_DR1} presents a large and relatively homogeneous spectroscopic dataset that is well suited for studies of low-ionization metal absorbers. With over a million quasar spectra in Data Release 1 (DR1), DESI expands the available sample size relative to previous surveys, enabling improved statistical measurements and more uniform selection across redshift \citep{Napolitano_2023,Anand_2025}. At the same time, the survey footprint does not cover the full sky and the dataset remains subject to observational selection effects related to the targeting strategy and varying observing conditions \citep{Chaussidon_2023}.

Cosmological simulations provide a framework for interpreting absorption-line measurements, linking observed statistics to processes such as gas accretion, outflows, and feedback \citep{Oppenheimer_2018, Hafen_2019, Peroux_2020, Nelson_2018, Nelson_2020, nelson2025_salsa}. By generating synthetic lines of sight, these models allow predictions for column density statistics and evolution, offering a crucial comparison for large observational samples. Such comparisons are essential for testing and refining theoretical models of galaxy formation.

With this in mind, the goals of this paper are threefold: (i) to quantify the distribution of Mg\,I, Mg\,II, and Fe\,II column densities in the IllustrisTNG simulation using a streamlined model that captures the essential physics, and to assess its robustness; (ii) to compare the resulting simulated column-density distributions with those inferred from the HIRES/UVES sample, as well as to confront the predicted Mg\,II incidence with existing statistical measurements; and (iii) to evaluate the extent to which DESI data can enable detailed comparisons with the simulation.

The paper is structured as follows. In \S\ref{sec:Churchill_catalog}, we present the HIRES/UVES sample used in this work. In \S\ref{sec:model}, we briefly introduce the IllustrisTNG simulation, describe the ionic abundance models, and outline the procedure for computing column densities. In \S\ref{sec:results}, we present the resulting column density distributions and cosmic incidence, and discuss the potential of DESI for further comparisons. Finally, we summarize this work in \S\ref{sec:conclusions}.

\section{Observational Absorption Line Data}
\label{sec:Churchill_catalog}

The primary sample of low-ionization absorbers used in this work is drawn from the catalog constructed by \citet{Churchill_2020}.
In the end of the paper, we also compare with the 
Mg\,II DESI catalog of \citet{Napolitano_2025}.

\subsection{HIRES/UVES Catalog}

\citet{Churchill_2020} performed a homogeneous Voigt profile analysis of archival high-resolution quasar spectra to characterize the kinematic structure of metal-line absorption systems. Their study is based on data obtained with HIRES on the Keck telescope and UVES on the Very Large Telescope (VLT). Both instruments have a spectral resolving power $R\approx 40000$–$50000$ (velocity resolution $\sim6$–$8\ \kms$) that allows for a detailed decomposition of complex absorption features into individual kinematic components.

The catalog contains 2989 entries, each corresponding to a unique kinematic component associated with 422 Mg\,II absorption systems spanning redshifts $0.19\leq z \leq 2.55$, detected along 249 quasar sightlines. For each component, the catalog provides measurements of the redshift $z$, column densities of various low-ionization metals including Mg\,I, Mg\,II and Fe\,II ($\rm N_{\rm Mg\,I},\ \rm N_{\rm Mg\,II},\ \rm N_{\rm Fe\,II}$) and the Doppler parameter $b$. The sample spans a wide range of absorption strengths, ranging from optically thin to highly saturated systems, with rest-frame EW for the Mg\,II $\lambda2796$ transition in the range $0.006\,\text{\AA} \leq \text{W}_0^{2796}\leq 6.23\,\text{\AA}$.

\citet{Churchill_2020} employed 
simultaneous Voigt profile fitting 
and the full instrumental resolution  
to resolve narrow kinematic components within each absorption system.
Weak systems typically exhibit only a small number of kinematic components, whereas stronger systems display significantly more complex velocity structures, reflected in many more components over broader velocity spreads. The resulting component-level column densities, derived in a uniform high-resolution framework, provide a robust basis for a comparison of their statistical properties with 
those of hydrodynamical simulations.

\subsection{EW of HIRES/UVES absorbers}
\label{sec:EW_HIRES}

To compare the observed incidence of Mg\,II absorbers with cosmological simulations we derive the EW $\rm W_0^{2796}$ by solving the Mg\,II Voigt-profile curve of growth using the catalog values of $\rm N_{Mg\ II}$ and $b$. We tabulate $\rm W_0^{2796}(N_{Mg\ II}, \it b)$ on a grid with spacings $\Delta \log \rm N_{Mg\ II} \approx 0.14\ cm^{-2}$ and $\Delta \log b \approx 0.07\kms $. For each absorption component, we generate 1000 random realizations of $\rm N_{Mg\ II}$ and $b$ drawn from their reported uncertainties to estimate $\rm W_0^{2796}$ and its associated uncertainty. Components for which either $\rm N_{Mg\ II}$ or $b$ is unconstrained were excluded from this procedure. 
\begin{figure}
    \centering
    \includegraphics[width=0.42\textwidth]{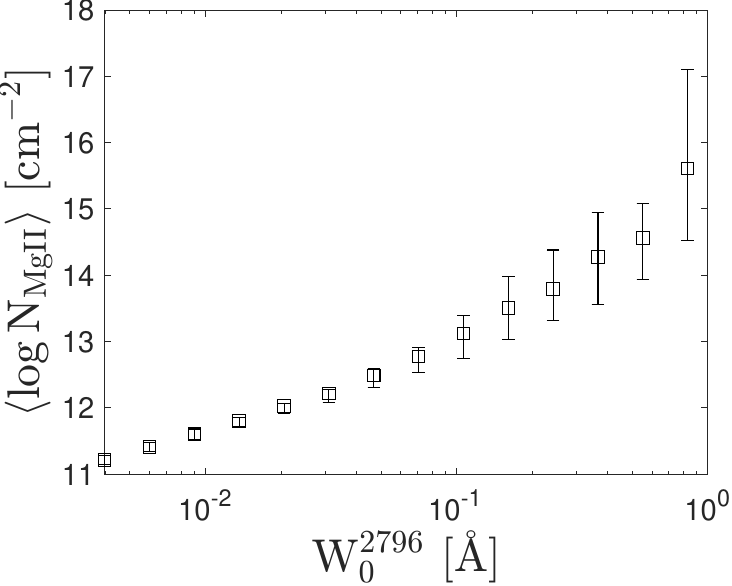}
    \caption{Average $\log \rm N_{Mg\,II}$ in bins of $\rm W_0^{2796}$ for the HIRES/UVES absorption components using $b$ values measured by \citet{Churchill_2020}.  
    }
    \label{fig:EW_N}
\end{figure}
 In Fig.~\ref{fig:EW_N} we present the resulting $\rm W_0^{2796}-N_{Mg\ II}$ relation, where $b$ is marginalized over. As $\rm W_0^{2796}$ increases, the characteristic column density also rises, accompanied by increasing scatter that reflects the growing complexity and saturation of stronger absorption systems.

\section{Simulated low-ionization metal absorbers}
\label{sec:model}

\subsection{The IllustrisTNG simulations}
\label{sec:TNG}

The IllustrisTNG project (hereafter TNG) is a series of comprehensive gravitational, magnetohydrodynamical simulations of galaxy formation \citep{TNG_DR, TNG_1, TNG_2, TNG_3, TNG_4, TNG_5, TNG_6, TNG_7}. In this paper, we use the high-resolution TNG50-1 simulation \citep{TNG50a, TNG50b}, which implements the following cosmology: $h=0.6774,\ \Omega_m=0.3089,\ \Omega_\Lambda=0.6911,\ \Omega_b=0.0486,\ \sigma_8=0.8159$ and $n_s=0.9667$. It evolves $2\times 2160^3$ dark matter and gas particles in a box of comoving side length $L=35\hmpc$, achieving a baryonic mass resolution $\sim 8\times 10^4\ M_\odot$. 
TNG has been shown to successfully reproduce key observational properties of both the CGM and IGM. In particular, it matches the observed column density and EW distributions of high-ionization species such as O\,VI and C\,IV \citep{Nelson_2018, nelson2025_salsa}, the Mg\,II covering fractions around luminous red galaxies at $z\sim0.5$ \citep{Nelson_2020}, as well as the statistical properties of the Ly$\alpha$ forest \citep{Khaire_2023}. Moreover, TNG accurately reproduces the X-ray emission from hot galactic halo gas across a wide range of redshifts and stellar masses \citep{Barnes_2018, Truong_2020}, and its predictions for the metal enrichment and thermodynamic structure of galactic and intergalactic gas are broadly consistent with observations \citep{Torrey_2019, Artale_2021}. 
Here, we employ TNG to study lower ionization absorption systems of Mg\,I, Mg\,II and Fe\,II and compare them with observations.

\subsection{Line Absorption Modelling in TNG}
\label{subsec:model}

TNG does not explicitly track the ionization states of individual elements. 
To model the abundances of Mg\,I, Mg\,II and Fe\,II, we compute their ionization balance under typical CGM and IGM conditions. For clarity, we present the ionization formalism below using Mg as the explicit example, with all equations written for the Mg ionic species. The same procedure is applied analogously to Fe by solving the corresponding ionization balance equations  using the respective Fe atomic rates.

For each gas cell, we compute the relative ionic abundances using an equilibrium ionization model that includes both collisional processes and photoionization in the presence of a uniform, redshift-dependent cosmic UVB. The resulting ionization fractions are then combined with the information provided by TNG on the gas mass and elemental abundances to obtain spatially resolved number densities for the ions of interest.

\subsection{Ionization Equilibrium Model}

We compute the relative Mg ionic abundances $\chi_{\text{Mg}}^{m} \equiv n(\text{Mg}^{m+})/\sum_m n(\text{Mg}^{m+})$ in each gas cell under the assumption of ionization equilibrium. The steady-state abundances are obtained by solving the linear system
\begin{equation}
     R^{m,m}\chi_{\text{Mg}}^{m}+R^{m,m-1}\chi_{\text{Mg}}^{m-1} + R^{m,m+1}\chi_{\text{Mg}}^{m+1} = 0 \;,
\end{equation}
where $R^{i,j}$ is a tridiagonal rate matrix describing transitions between adjacent ionization states. The diagonal elements $R^{m,m}$ encode the net loss rate of charge state $m$ by ionization to a higher state or radiative recombination to a lower state, while the off-diagonal elements ($R^{m,m-1}$, $R^{m,m+1}$) encode the population rates of state $m$ by ionization (from below) or recombination (from above). 
The expressions of the rates are given by
\begin{align}
\label{eq:ion_eq_rates}
    & R^{m,m} = -(n_e \alpha_{\text{RR}}^{m} + n_e \alpha_{\text{CI}}^{m} + \Gamma_{\rm{PI}}^{m}) \\ & R^{m,m-1} = n_e \alpha_{\text{CI}}^{m-1} + \Gamma_{\rm{PI}}^{m-1} \nonumber \\ & R^{m,m+1} = n_e \alpha_{\text{RR}}^{m+1} \nonumber \;,
\end{align}
where $n_e$ is the free electron density, and the rate coefficients $\alpha_{\text{CI}}^{m}(T)$, $\alpha_{\text{RR}}^{m}(T)$ for collisional ionization and radiative recombination of the ion $\text{Mg}^{m+}$ depend only on the temperature. Furthermore, $\Gamma_{\rm{PI}}^{m}$ is the non-collisional photoionization rate of $\text{Mg}^{m+}$ in the presence of a radiation field. 

A non-trivial solution for $\chi_{\text{Mg}}^{m}$ is obtained upon fixing the normalization $\sum_{m}\chi_{\text{Mg}}^{m}\equiv 1$.
We note that in the absence of a radiation field ($\Gamma_{\rm{PI}}^{m} \equiv 0$), the steady-state ionic abundances depend solely on the gas temperature since, in this case, all the rates in Eq.~\eqref{eq:ion_eq_rates} linearly scale with $n_e$, which factorizes out. When photoionization is included however, the steady-state abundances are generally functions of $n_e$ and $T$.

\subsection{Implementation in $\rm{TNG}$}

For the rate coefficients $\alpha_{\text{CI}}^{m}(T)$ and $\alpha_{\text{RR}}^{m}(T)$, we use functions provided in the \texttt{CHIANTI} atomic database \citep{Chianti_1,Chianti_2}. The photoionization rates are computed from the background radiation field with a photon frequency-dependent specific intensity $J_\nu$ and photoionization cross sections $\sigma_{\rm{PI}}^{m}(\nu)$ as
\begin{equation}
    \Gamma_{\rm{PI}}^{m} = 4\pi \int \frac{J_\nu}{h\nu}\sigma_{\rm{PI}}^{m}(\nu) d\nu \;,
\end{equation}
where $h$ is the Planck constant. In TNG, gas is evolved under the influence of the \citet{Faucher_Giguere_2009} cosmic UVB, using the 2011 update to that model. To ensure consistency, we adopt the same $J_\nu$ when computing the Mg ionization fractions $\chi_{\text{Mg}}^{m}$. Furthermore, we use the analytic fits of \citet{Verner_1996} for the Mg and Fe photoionization cross sections. Typical values for Mg are $\Gamma_{\rm{PI}}^{0}\sim 10^{-12}\ \rm{s^{-1}}$ and $\Gamma_{\rm{PI}}^{1}\sim 10^{-14}\ \rm{s^{-1}}$ in the redshift range $1\lesssim z \lesssim 2$. 
Throughout this work, we consider two variants of the ionic abundance model: 
\begin{itemize}
\item Coll, which includes only collisional processes (no radiation field).
\item Coll+UVB, which additionally includes the cosmic UVB.
\end{itemize}
Using the ionization and recombination rate coefficients, we produce gridded solutions for $\chi_{\text{Mg}}^{0}(n_e,T)$, $\chi_{\text{Mg}}^{1}(n_e,T)$ and $\chi_{\text{Fe}}^{1}(n_e,T)$ at redshifts $z=0.7$, 1, 1.5 and 2. We use these grids to determine the relative abundance of the ions in each gas cell of TNG.

\begin{figure}
    \centering
    \includegraphics[width=0.45\textwidth]{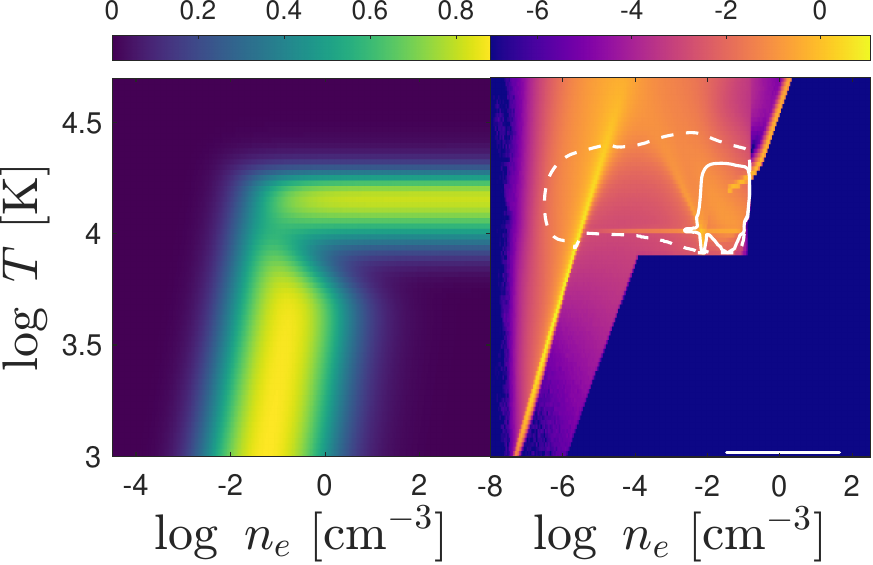}
    \caption{\textbf{Left: }Steady state fractional abundance of Mg\,II $(\chi_{\text{Mg}}^{\rm{1}})$  as a function of $n_e, T$, including the 
    UVB field at $z=1$. The horizontal branch above $T=10^4$\,K is where collisional ionization and radiative recombination dominate (both scale with $n_e$), whereas at lower temperatures UVB photo-ionization balanced by radiative recombination determines the Mg\,II population, which peaks around $n_e\approx 0.1$\,cm$^{-3}$. 
    \textbf{Right: } TNG gas-mass distribution at $z=1$ (plotted as a normalized PDF).  
    Solid and dashed white contours enclose 99\% of the Mg\,II mass for the Coll+UVB and the Coll models, respectively. 
    The white line at $T=10^3\Kel$ is part of the Coll+UVB contour. 
    It marks the contribution from star forming cells, whose temperature is not properly modeled by TNG and was uniformly set to $T=10^3\Kel$.} 
    \label{fig:MgII_abund_grid}
\end{figure}

The left panel of Fig.~\ref{fig:MgII_abund_grid} shows the solution for the Mg\,II abundance in the $(n_e,T)$ plane for the Coll+UVB model at $z=1$.
The primary effect of the radiation field is to sustain Mg\,II at temperatures lower than $\sim 2\times 10^4\Kel$, provided that the electron density is sufficiently low such that collisional processes do not dominate ($n_e\lesssim 0.1\ \rm{cm^{-3}}$).
Following the modelling of Mg\,II emission by \citet{Nelson_2021}, we manually set the temperature of all star forming gas cells to $T=10^3\Kel$, to match the temperature of the cold phase gas in the two-phased model used by TNG. 
The ionic masses in the $i$th gas cell are computed as 
\begin{align}
\label{eq:cell_ionic_abund}
     & m_{\text{Mg\,I}}^{i} = m_{\text{gas}}^{i}\text{X}_{\text{Mg}}^{i}\chi_{\text{Mg}}^{0,i} \\ & m_{\text{Mg\,II}}^{i} = m_{\text{gas}}^{i}\text{X}_{\text{Mg}}^{i}\chi_{\text{Mg}}^{1,i}\nonumber \\ & m^i_{\text{Fe\,II}} = m_{\text{gas}}^{i}\text{X}_{\text{Fe}}^{i}\chi_{\text{Fe}}^{1,i} \;, \nonumber
\end{align}
where $m_{\text{gas}}^{i}$ is the cell total gas mass and $\text{X}_{\text{Mg}}^{i}$, $\text{X}_{\text{Fe}}^{i}$ denote the elemental abundances (mass fractions) of Mg and Fe in the cell. The right panel of Fig.~\ref{fig:MgII_abund_grid} shows the distribution of TNG gas cells in the $(n_e,T)$ plane. The white contours indicate where the Mg\,II mass is concentrated in the two models. 
In the Coll+UVB model $\sim 78\%$ of the total Mg\,II mass is in star forming cells, and the model results in $\sim 2.1$ times more Mg\,II mass than in the Coll model. However, when excluding these cells, the Coll model produces $\sim 2.2$ times more Mg\,II, as there is no UVB to further ionize Mg beyond Mg\,II.

\begin{figure}
    \centering
    \includegraphics[width=0.4\textwidth]{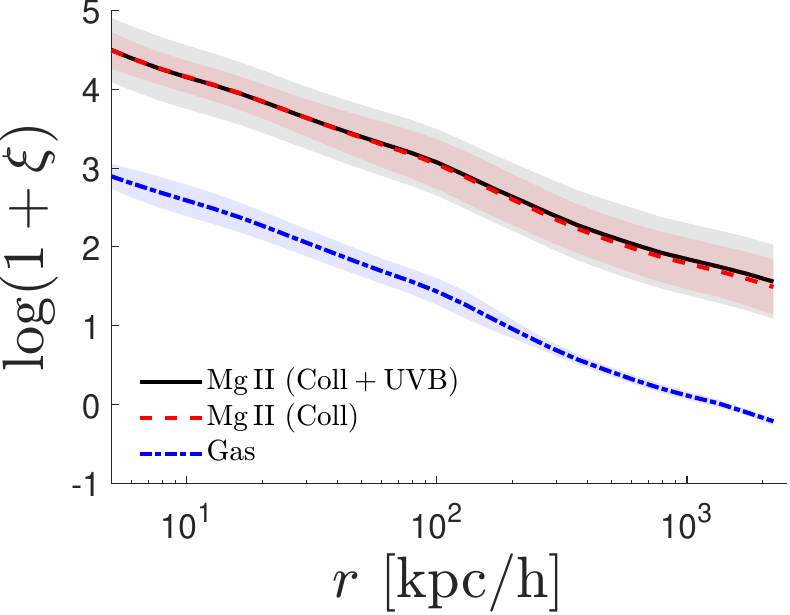}
    \caption{Three-dimensional two-point autocorrelation functions $\xi(r)$ of Mg\,II and the total gas distributions at $z=1$, measured from TNG as a function of comoving separation $r$. The Mg\,II mass exhibits a much stronger autocorrelation, consistent with its association with dense regions.}
    \label{fig:TNG_MgII_2PCF}
\end{figure}

To better understand the spatial distribution of ions predicted by the models, we show in Fig.~\ref{fig:TNG_MgII_2PCF} the three-dimensional, two-point autocorrelation function of Mg\,II at $z=1$, compared to that of the total gas distribution. The correlation functions and their uncertainties are computed using the Landy–Szalay estimator, based on samples of $5\times10^6$ gas cells randomly drawn from four subregions of the simulation box, each with comoving volume $(5\hmpc )^3$, and each cell is weighted by $m_{\text{gas}}^{i}$ or  $m_{\text{Mg\,II}}^{i}$ (of the corresponding model). Across the mildly-to-strongly sampled nonlinear scales, Mg\,II shows a markedly higher clustering amplitude than the overall gas, reflecting its preferential association with dense environments, with little difference between the Coll or Coll+UVB models. Notably, Mg\,II remains well correlated beyond $r\approx 1200 \hkpc$, where the gas correlation turns negative. The weak change in the scale dependence around $r\sim 200 \hkpc$ likely corresponds to the virial radius of typical virializing halos at $z=1$ (with mass $M_\ast\approx 10^{11} \hmsun$), marking the transition between the one-halo and two-halo clustering regimes.

\subsection{Simulated Column Densities}
\label{sec:TNG_MgII_abundances_grid}

The final step in constructing the column density field is to project the discrete TNG gas cells onto a two-dimensional grid using a smoothed-particle hydrodynamics (SPH) kernel. In this framework, each grid point corresponds to a unique line of sight through the simulation volume. For each TNG gas cell, we compute the total number of ions by dividing the cell’s ionic mass (Eq.~\eqref{eq:cell_ionic_abund}) by the atomic mass of the corresponding element (24.1 and 55.4 proton masses for Mg and Fe, respectively). The column densities are computed by projecting the particle numbers onto a two-dimensional grid using a standard cubic-spline SPH kernel \citep{Monaghan_1992}, where each particle is smoothed over two layers of grid cells ($R=2$), corresponding to $(2R+1)^2=25$ pixels on the square grid. The kernel size of each gas cell, $h_i$ (in pixel units), is derived from its comoving volume $V_i$ and the comoving grid resolution $\Delta x = L/N_{\text{grid}}$ as
\begin{equation}
    h_i=V_i^{1/3}/\Delta x \;.
\end{equation}
The grid size $N_{\text{grid}}$ is chosen separately for each model such that the ionic-mass-weighted mean kernel size satisfies $\langle h_i\rangle \approx R$, ensuring balanced smoothing. Since the gas distribution differs between the models, the average kernel size also differs. In the Coll+UVB model, ionic mass is more concentrated in denser gas with smaller kernel sizes, we therefore adopt $70000\times70000$ grids, while for Coll we satisfy $\langle h_i\rangle \approx R$ by taking coarser $25000\times25000$ grids, since there the distribution shifts toward more diffuse CGM gas with bigger cell volumes. A small fraction of cells have $h_i > R$ by factors of a few, leading to partial truncation of their ionic masses. These correspond to the most dilute cells and contribute negligibly to the total line-of-sight column density. Overall, the Coll+UVB SPH projections retain more than 99\% of the total input ionic mass, and over 90\% for Coll. The robustness of these results is examined further at the end of this section.

\begin{figure}
    \centering
    \includegraphics[width=0.45\textwidth]{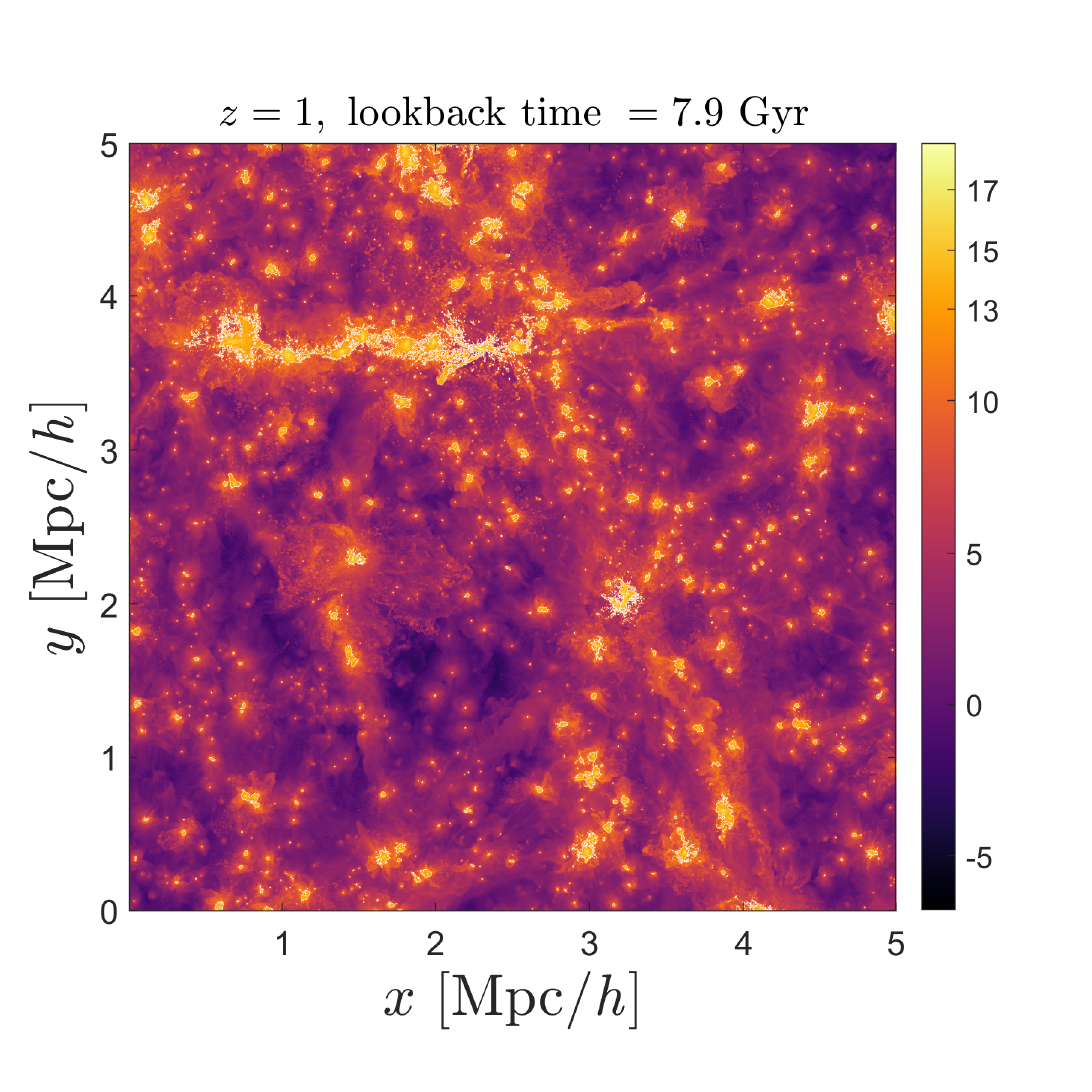}
    
    \caption{A comoving $5\times 5\ [h^{-1} \rm{Mpc}]^2$ cutout of the gridded column density field computed in TNG50-1 at $z=1$, Coll+UVB model. 
    Colors show the column density of each line of sight $\log \rm{N_{Mg\,II}}\ [\rm{cm^{-2}}]$ as quoted by the colorbar. Thin white lines show contours of lines of sight with $\rm{N_{Mg\,II}} \approx 10^{12}\ \rm{cm^{-2}}$. }
    \label{fig:gridded_MgII_cd_vis}
\end{figure}

In Fig.~\ref{fig:gridded_MgII_cd_vis} we visualize a cutout of the Mg\,II column density field computed for the Coll+UVB model in TNG at $z=1$, using a smaller grid of $20000\times 20000$. The column densities span almost 25 orders of magnitude. We find that absorption features typically detectable by HIRES/UVES ($\rm{N_{Mg\,II} \gtrsim 10^{12}\ cm^{-2}}$) largely correspond to the CGM of galaxies, or IGM in dense clusters, indicating that Mg\,II absorbers are discrete, localized objects, with similar behavior for Mg\,I and Fe\,II absorbers. 

Since the incidence of these high-column-density absorbers along sight lines through TNG is low, we will assume that each line of sight intersects exactly one such absorber when the total sight line column density is $\rm{N}\geq 10^{11}\ cm^{-2}$. Concretely, we compute the gridded column densities once in the $xy$ plane of the TNG snapshots at $z=0.7$, 1, 1.5 and 2.
This yields a total of $N_{\text{grid}}^2 = 4.9\times 10^9$ (Coll+UVB) and $N_{\text{grid}}^2 = 6.25\times 10^8$ (Coll) unique lines of sight of fixed length that we use to sample the column density PDF of the ions, $\phi(\rm{N_{Mg\,I}}),\ \phi(\rm{N_{Mg\,II}})$ and $\phi(\rm{N_{Fe\,II}})$. In practice, only $\approx 2.8\%$ of lines of sight have $\rm{N_{Mg\,II}}\geq 10^{12}\ cm^{-2}$ at $z=1$.

\begin{figure}
    \centering
    \includegraphics[width=0.45\textwidth]{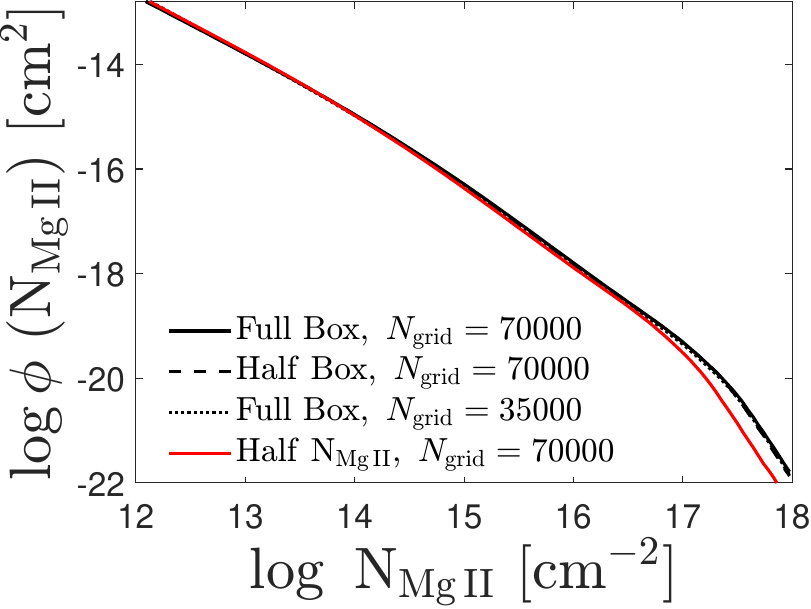}
    \caption{Systemic examination of the Mg\,II column density PDF, normalized to $10^{12} \leq\rm{N_{Mg\,II}}\leq 10^{18}\ cm^{-2}$. The solid line shows results for the $xy$ plane of TNG50-1 at redshift $z=1$ with grid size $N_{\text{grid}}=70000$, integration over full $z$ axis. The red line shows the same results but when all column densities are multiplied by $0.5$. The dashed line shows $N_{\text{grid}}=70000$ again but when only the upper half of the box is integrated ($z\geq L/2$). Finally, the dotted line shows full box integration but with worsened spatial resolution $N_{\text{grid}}=35000$. }
    \label{fig:MgII_cd_sim_sys}
\end{figure}

To assess the robustness of the simulated column density PDFs of strong absorption systems, we perform several tests on $\phi(\rm{N_{Mg\,II}})$, shown in Fig.~\ref{fig:MgII_cd_sim_sys}. The fiducial column density PDF (solid black) is compared to: 1) a PDF when only half of the box is integrated (dashed black), 2) a PDF where the spatial resolution is degraded from $\Delta x=0.5 \hkpc$ to $\Delta x=1 \hkpc$ (dotted black), and 3) a PDF when all the column densities of the fiducial PDF are multiplied by $1/2$ (solid red). We can see that the first three PDFs are almost indistinguishable from one another all the way up to the highest column densities, and they notably differ from the half column density case, particularly at $\rm{N_{Mg\,II}\gtrsim 10^{16.5}\ cm^{-2}}$ where the PDF deviates from a power law. These results strengthen the assertion that the absorbers are discrete objects and show that the column density statistics are convergent in terms of grid resolution. This enables a straightforward and meaningful comparison between PDFs in simulations and in observations. 

\section{Results}
\label{sec:results}
\subsection{PDF comparison}
\label{sec:pdf_comparison}
Observations suffer from bias associated with incompleteness at low column densities. To mitigate it, we normalize the column density PDFs of both TNG and the observations over the fixed ranges
\begin{align}
    & \rm{N_{Mg\,I}\in (10^{11.4},10^{13})}\ cm^{-2} \\ & \rm{N_{Mg\,II}\in (10^{12.4},10^{17})}\ cm^{-2} \nonumber \\ & \rm{N_{Fe\,II}\in (10^{12.6},10^{17})}\ cm^{-2} \nonumber \;, 
\end{align}
within which the measurements are nearly complete \citep{Churchill_2020}.

To construct the comparative column density PDFs, we first partition the absorbers into four redshift bins, $z\in [0.55, 0.85]$, $[0.85,1.15]$, $[1.35, 1.65]$, $[1.85, 2.15]$ 
selected to bracket the TNG redshifts. Furthermore, we exclude observed absorbers for which either the column density or the Doppler parameter $b$ is unconstrained. 
For each
absorption component, we use the reported column density and its associated uncertainty 
to generate 5000 Monte Carlo realizations of the column density, assuming a normal distribution. These realizations are used to estimate the column density PDF and its associated uncertainty. To plot the PDFs, we adopt the fixed column density ranges described above, with logarithmic bin widths of $\Delta \log \text{N} = 0.2$ (N in $\rm{cm^{-2}}$). From the ensemble of realizations, we compute the median PDF and its 95\% confidence interval.
\begin{figure}
    \centering
\includegraphics[width=0.45\textwidth]{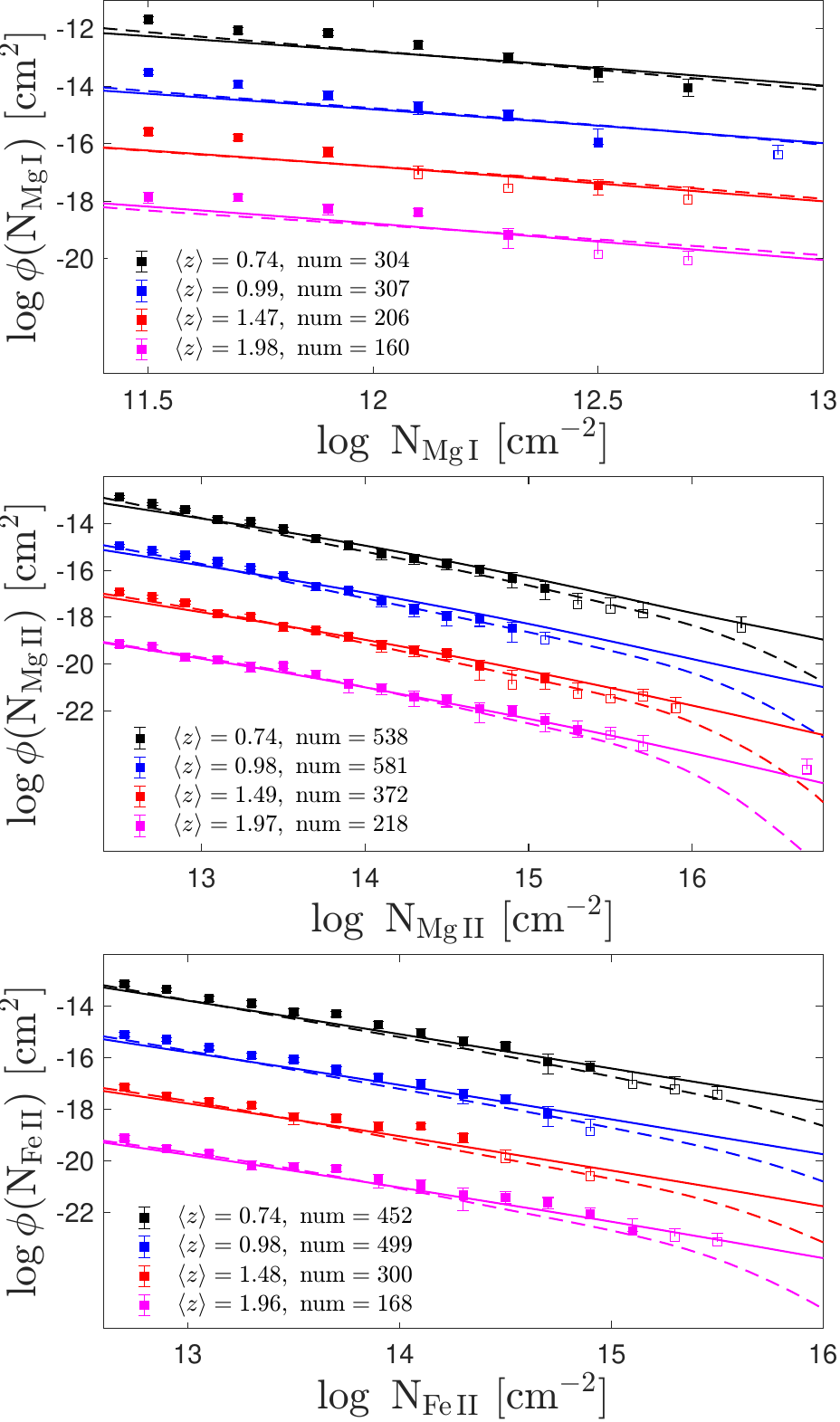}
    
    \caption{Column density PDFs for the three ions examined in the paper. Each panel shows
    four redshift bins (see Sec \S \ref{sec:pdf_comparison}). Squares represent results from the HIRES/UVES catalog; the insets indicate the average redshift $\langle z \rangle$ and the total number of absorbers (num) used to compute the PDFs. Open squares denote 
    upper limits. 
    Solid and dashed lines show results for the Coll+UVB and Coll models, respectively. 
    Since the data and TNG PDFs in each redshift bin are similar, 
    redshifts $z\sim 1,\ 1.5,\ 2$ are vertically shifted by $-2, -4,$ and $-6$, respectively. }
    \label{fig:PDF_comparison_HIRES}
\end{figure}

The resulting PDFs are shown in Fig.~\ref{fig:PDF_comparison_HIRES}. For the three ions considered here the 
PDFs exhibit little redshift evolution between $0.7 \lesssim z \lesssim 2$, suggesting that individual metal absorption systems are already well established and do not evolve after they form \citep[c.f.][]{wu/etal:2025}. 
For this reason, we vertically shift the plots in Fig.~\ref{fig:PDF_comparison_HIRES} for clarity. 

TNG successfully reproduces the shape of the PDFs across almost the entire range of column densities probed. The two ionization models produce nearly the same results for optically thin to moderately saturated absorbers up to $\text{N}\lesssim 10^{14}\ \rm{cm^{-2}}$. At higher column-densities the models diverge. 
In the Coll+UVB model, the column density PDFs generally maintain a continuous slope toward higher decades in column density, whereas the Coll model exhibits a pronounced break at $\text{N} \sim 10^{15.5}\ \rm{cm}^{-2}$, beyond which the occurrence of denser absorbers drops rapidly. 
This effect is due to our assumption that star forming regions are at $T = 10^3 \Kel$ where collisions are totally ineffective (see Fig.\,\ref{fig:MgII_abund_grid}).
For this reason, at higher column densities, the PDFs for Mg\,II and Fe\,II hint at a preference for the Coll+UVB model. 
However, this distinction remains inconclusive, as nearly all measurements at $\rm{N} \gtrsim 10^{15}\ \rm{cm}^{-2}$ have only upper limits. 

Among the three ions, Mg I exhibits the poorest agreement with the data, with TNG predicting an incompatibly shallower slope. This discrepancy is largely expected for all neutral ions, as the multiphase ISM model implemented in TNG \citep{Springel_2003} is not designed to accurately resolve gas at temperatures below $T \lesssim 10^4\Kel$.

\subsection{Cosmic incidence $d\mathcal{N}/{dz}$}
\label{sec:cosmic_incidence}
\begin{figure*}[htp]
    \centering
    \includegraphics[width=0.8\textwidth]{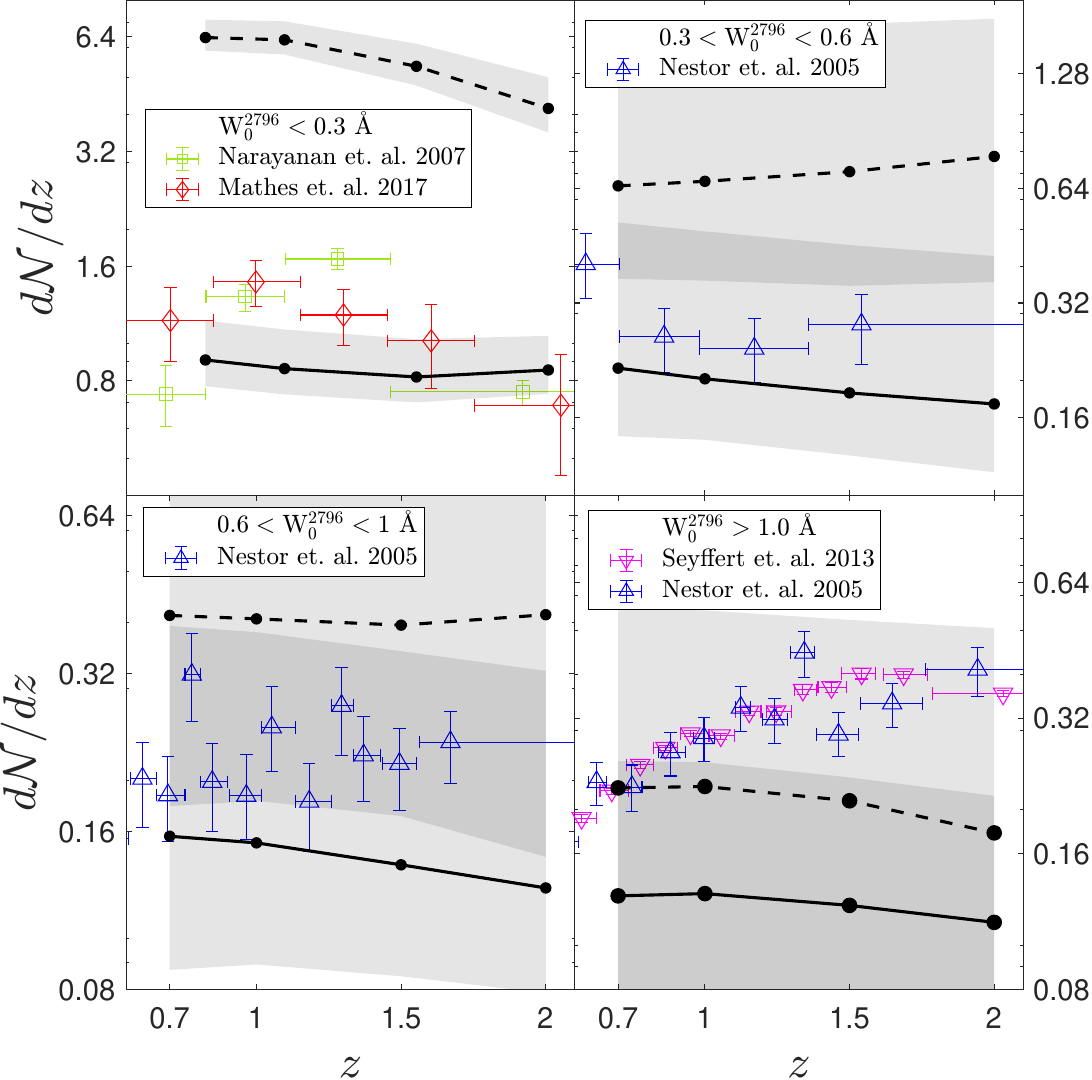}
    \caption{Comparison of the measured Mg\,II absorber incidence with TNG. Each panel shows the cosmic incidence in a different EW bin as quoted, and these are compared to TNG when mapped into the approximate corresponding column density bins (see \S\ref{sec:cosmic_incidence}), with the solid black lines and dashed lines corresponding to the Coll+UVB, Coll models respectively. The grey shades represent the model uncertainties, due to the adopted EW-column density mapping (see text), and dark grey is where they overlap.} 
    \label{fig:MgII_cosmic_incidence}
\end{figure*}
While the PDF captures the relative distribution of column densities, it hides crucial information about the absolute incidence of absorbers. In this section, we examine the cosmic incidence of Mg\,II absorbers as an additional test of TNG and the ionic abundance models. To make the comparison, we draw on earlier measurements of the cosmic incidence of Mg\,II absorbers reported in EW bins. These include weak to moderately saturated systems ($\mathrm{W_0^{2796}}<0.3\ \text{\AA}$) measured from VLT/UVES quasar spectra \citep{Narayanan_2007,Mathes_2017}, as well as more strongly saturated absorbers ($0.3<\mathrm{W_0^{2796}}<0.6\ \text{\AA}$, $0.6<\mathrm{W_0^{2796}}<1\ \text{\AA}$, and $\mathrm{W_0^{2796}}>1\ \text{\AA}$) derived from SDSS quasar samples \citep{Nestor_2005,Seyffert_2013}. Because direct EW statistics cannot be computed from TNG without modeling the turbulent broadening, we instead use the marginalized $\mathrm{W_0^{2796}}-\rm N_{Mg\,II}$ relation (see Sec.\,\ref{sec:EW_HIRES}) to approximately map these EW intervals onto four Mg\,II column-density bins: $\rm{N_{Mg\,II}}<10^{13.7^{+0.7}_{-0.3}},\ \rm{ N_{Mg\,II}}\in \left(10^{13.7^{+0.7}_{-0.3}},10^{14.3^{+0.8}_{-0.4}}\right),\ \rm{ N_{Mg\,II}}\in \left(10^{14.3^{+0.8}_{-0.4}},10^{15.0^{+1.2}_{-0.6}}\right)$ and $\rm{ N_{Mg\,II}}>10^{15.0^{+1.2}_{-0.6}},\ \rm{cm^{-2}}$. In TNG, the cosmic incidence is computed as the number of sightlines within a given column-density bin ($N_{\rm{abs}}$), divided by the total number of sightlines and by the redshift path length $\Delta z$,
\begin{equation}
    \frac{d\mathcal{N}}{dz} = \frac{N_{\text{abs}}}{N_{\text{grid}}^2 \Delta z} \;,
\end{equation}
where $\Delta z$ is computed from the comoving distance-redshift relation,  using the TNG side length and cosmology. For the lowest EW bin, the inferred incidence is sensitive to the adopted lower column density threshold for detecting the absorber. Since \citet{Narayanan_2007} report 86\% completeness down to $\rm{W_0^{2796}=0.02\ \text{\AA}}$, we additionally impose $\rm{N_{Mg\,II}} \geq  10^{12}\ cm^{-2}$ using the relation of Fig.~\ref{fig:EW_N} again. 
Adopting instead a lower threshold of $\rm{N_{Mg\,II}} \geq  10^{11.7}\ cm^{-2}$ would increase the inferred incidence in both models by only $\approx 10-15\%$.

The results are shown in Fig.~\ref{fig:MgII_cosmic_incidence}. At $\mathrm{W_0^{2796}<0.3\ \text{\AA}}$ and $0.3<\mathrm{W_0^{2796}}<0.6\ \text{\AA}$, where most Mg\,II absorbers reside, the observed cosmic incidence is reasonably consistent with the Coll+UVB model. The Coll only model overestimates the number of Mg\,II systems likely since the effect of (over)photoionization is not included.
At higher EWs, in the range $0.6<\mathrm{W_0^{2796}}<1\ \text{\AA}$, the observational measurements lie between the two models and are consistent with both (dark grey area in the figure).

For the strongest absorbers, $\mathrm{W_0^{2796}}>1\ \text{\AA}$, both models do not produce enough Mg\,II absoption systems, although the Coll model is formally consistent with the data within its uncertainties.
We can think of two explanations for this discrepancy. One is that these systems occur in cold star forming regions, where TNG fails to produce their correct temperatures and our assumption of a uniform $T = 10^3$\Kel\ is an oversimplification. The second reason may be the way we convert measured column densities to EWs (Sec.\,\ref{sec:EW_HIRES}), while marginalizing over the $b$ parameter distribution. If the high-EW systems have unusually high turbulence ($b$) \citep[e.g.,][]{Kakoly_2025}
our conversion to EWs would underestimate their true values.
Finally, both models fail to capture the observed increase in incidence of high-EW systems towards $z\sim2$ \citep{Nestor_2005, Seyffert_2013}. Interestingly, \citet{nelson2025_salsa}, who performed a detailed Mg\,II spectral synthesis by ray-tracing lines of sight through TNG100-1, also find that simulations do not account for all observed absorbers with $\rm W_0^{2796}\gtrsim 0.6\ \text{\AA}$. 
High spatial resolution in the CGM 
reveals pockets of cool gas that blend out at lower resolution \citep[e.g.][]{Hummels_2019,van_de_Voort_2019,Lucchini2026}, and may also be the culprit for this discrepancy, although its appearance only at high $\mathrm{W_0^{2796}}$ and at $z\sim 2$ is unclear.
Finally, \citet{Grauer2023} computed a shallow increase of X-ray opacity in TNG between $1 < z < 2$ that stops short of explaining the observed opacity.

We conclude that TNG may be underestimating the metal content outside of galaxies, and that cosmic incidence evolution is a better way to distinguish between models than the column density PDF, which seems to be agnostic to redshift (Sec.\,\ref{sec:pdf_comparison}).

\subsection{Comparison with DESI}
\label{sec:DESI}
\begin{figure}
    \centering
\includegraphics[width=0.45\textwidth]{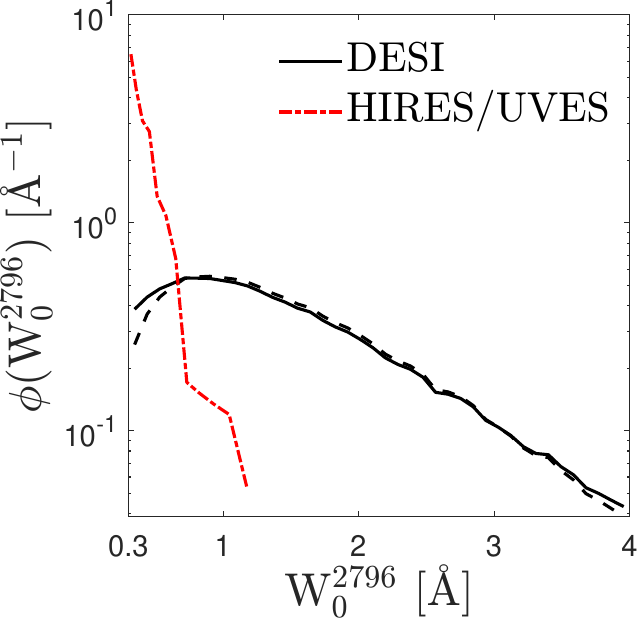}
    
    \caption{PDFs of the rest-frame EW of the Mg\,II $\lambda2796$ line for absorbers in the redshift range $0.5<z<2$, normalized over $0.3<\rm W_0^{2796}<4\ \text{\AA}$. The black solid and dashed curves respectively show the DESI distributions with and without a completeness correction, while the dotted red curve corresponds to the HIRES/UVES sample. The two PDFs are seen to be totally different.}
    \label{fig:EW_PDF}
\end{figure}
\begin{figure*}[htp]
    \centering
    \includegraphics[width=0.8\textwidth]{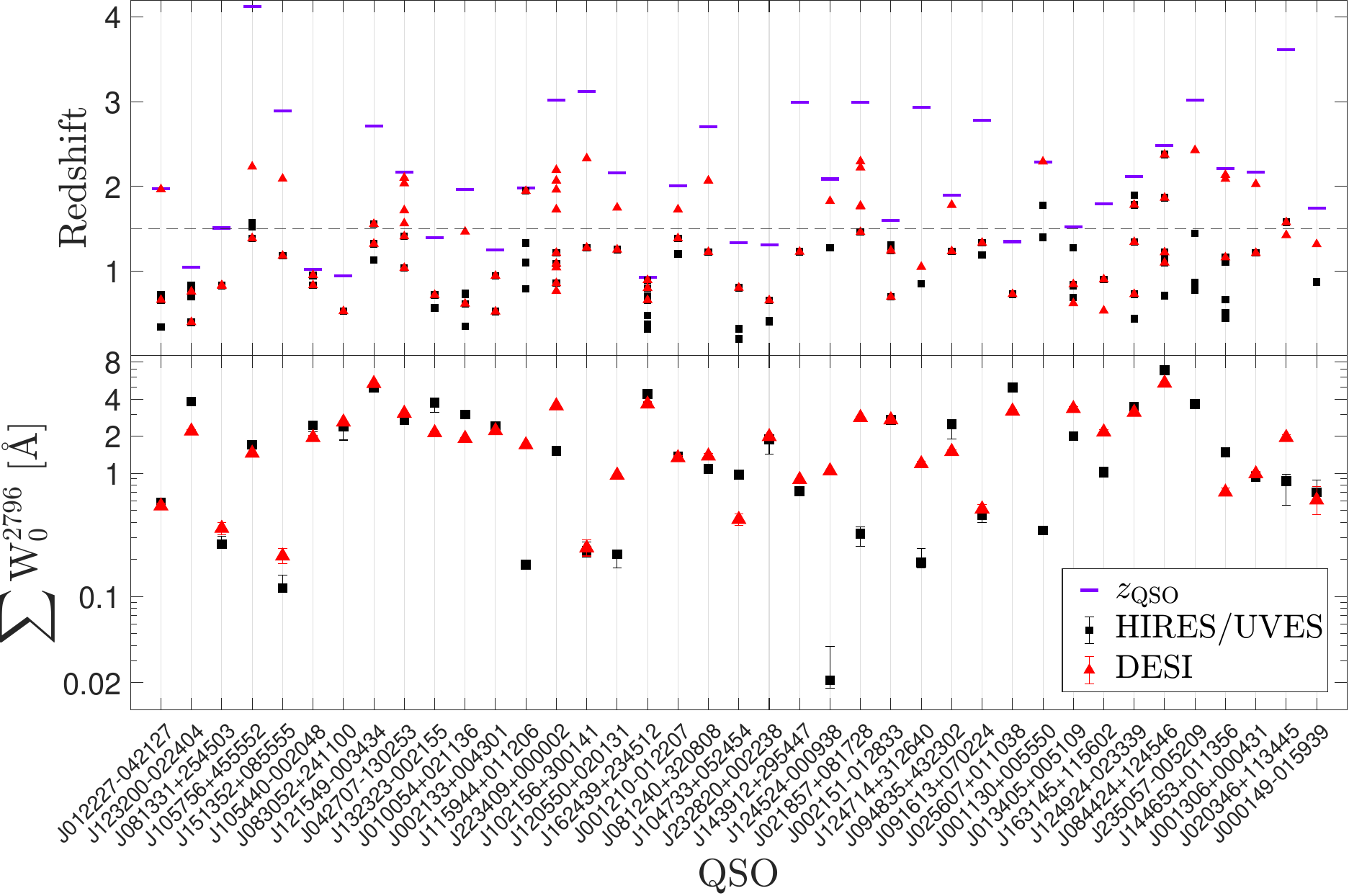}
    \caption{Mg\,II absorption properties for the identified 39 quasars common to both catalogs. Quasar names are given in J2000 format. \textbf{Top panel}: Redshifts of individual absorption components/systems identified in each catalog (black squares for HIRES/UVES, red triangles for DESI). Error bars for redshifts reported in the catalogs are omitted for clarity, as they are typically small (relative uncertainties $\lesssim 1\%$). Purple lines indicate the quasar redshifts (as fitted by DESI), and the thin black dashed line marks $z=1.5$. \textbf{Bottom panel}: Total rest-frame equivalent width, $\sum \rm W_0^{2796}$, along each quasar sightline, including only absorbers with $z \leq 2$.} 
    \label{fig:overlapping_quasars}
\end{figure*}
As part of DESI's extensive spectroscopic observations,  \citet{Napolitano_2023,Napolitano_2025} developed methods to detect Mg\,II absorbers in DESI quasar spectra, producing a value-added catalog (VAC) from the Early Data Release and DR1. The DR1 catalog contains over 270000 systems at $0.3\lesssim z\lesssim 2.5$, with the total expected to reach 800000 in future releases. This subsection aims to assess the feasibility of comparing DESI Mg\,II observations with TNG.

In Fig.~\ref{fig:EW_PDF}, we show the EW distributions $\phi(\rm W_0^{2796})$ for the DESI Mg\,II catalog and for the HIRES/UVES sample (recovered by inverting the curve of growth, see Sec.\,\ref{sec:EW_HIRES}). For a consistent comparison, both catalogs are restricted to the redshift range $0.5<z<2$
and to EWs $0.3 < \rm W_0^{2796} < 4\ \text{\AA}$, where a completeness grid is available for DESI. The DESI sample exhibits a systematic excess of high-EW systems relative to HIRES/UVES. 
We ascribe this difference primarily to the lower spectral resolution of DESI, which results in blending of adjacent narrow absorption components.
As a result, several low-EW components are counted as a large one. An additional factor that may affect the DESI PDF at low EW is the sampling of the DESI completeness grid. The correction provided by \citet{Napolitano_2023} is tabulated on a relatively coarse grid in EW, which may limit the accuracy of the completeness estimate near the lowest EWs considered here. A finer sampling of the completeness grid could potentially increase the corrected number of weak absorbers and partially modify the shape of the DESI EW distribution at low EW.

To further assess the consistency between the catalogs, we identified 39 quasars common to both samples. The top panel of Fig.~\ref{fig:overlapping_quasars} shows the redshifts of all absorbers detected along each quasar sightline. DESI identifies substantially fewer distinct absorption components than HIRES/UVES and exhibits an excess of absorbers at $z \gtrsim 1.5$ with no counterparts in the high-resolution sample. 
Additionally, the bottom panel of Fig.~\ref{fig:overlapping_quasars} presents the total $\rm W_0^{2796}$ summed over each quasar sightline for all absorbers with $z \leq 2$, where both instruments are sensitive. This integrated EW should be conserved irrespective of spectral resolution, even if individual sub-components are blended in the DESI spectra. For the majority of quasars, the total EW reported by the two catalogs is in reasonable agreement; however, a small subset shows discrepancies. These are tentatively a result of mis-identification of Mg\,II lines.

Taken together, these results suggest that while DESI is effective in identifying the presence of Mg\,II absorption along quasar sightlines in large numbers, line blending due to its lower spectral resolution limits the utility of the catalog for direct comparisons with TNG.
Detailed spectral synthesis performed on TNG to match DESI’s resolution and sensitivity, and improved line-identification procedures might mitigate DESI's disadvantage, but are beyond the scope of the present work.
\section{conclusions}
\label{sec:conclusions}
This paper presents a simplified yet physically motivated model for the spatially resolved distributions of Mg\,II, Mg\,I, and Fe\,II that can be applied to hydrodynamical simulations of galaxy formation. Based on the IllustrisTNG TNG50-1 simulation, we considered two ionization prescriptions: a purely collisional model (Coll), and a model that additionally includes a uniform ionizing ultraviolet background (Coll+UVB). In both cases, ionic abundances were computed under the assumption of steady-state ionization, and tabulated on grids spanning the relevant temperature and density parameter space. These precomputed solutions were combined with gas properties from TNG to assign ionic masses to all gas cells. The resulting ionic mass distributions were then projected onto two-dimensional grids using an SPH-based scheme, enabling the construction of two observable statistics directly comparable to data: the column density probability distribution function, $\log \phi(\rm N)$, and for Mg\,II (only) cosmic incidence $d\mathcal{N}/dz$. Our results can be summarized as follows:
\begin{itemize}
    \item TNG correctly produces the HIRES/UVES column density PDFs above $\rm N \gtrsim  10^{11.4}\ cm^{-2}$ with both ionization prescriptions. For Mg\,I (only) TNG produces a shallower slope than observed in the data; we attribute this discrepancy to the fact that the multiphase ISM model of TNG is not intended to model gas below $T \lesssim 10^4 \Kel$. 
    
    \item The column density PDFs show little evolution with redshift between $0.7<z<2$, particularly for low column densities, suggesting that many of these absorption systems formed at early times and did not significantly evolve during this epoch.

    \item The Mg\,II cosmic incidence $d\mathcal{N}/{dz}$ analysis shows a strong preference for the Coll+UVB model. The preference is particularly strong for the weakest absorbers ($\rm W_0^{2796} < 0.3\ \text{\AA}$), which dominate the population of Mg\,II systems, but also for $0.3<\rm W_0^{2796} < 0.6\ \text{\AA}$, indicating that such absorbers correspond to low-density, photoionized gas typically at the outer CGM. At higher EWs neither model correctly produces $d\mathcal{N}/{dz}$, and in particular TNG fails to capture its rise towards $z \sim 2$.
    
    \item Consequently, accurate modeling of absorber cosmic incidence provides a more effective diagnostic of ionization prescriptions than comparisons based on column density PDFs.
    
    \item Analysis of the DESI Mg\,II absorber catalog and overlapping quasars with HIRES/UVES shows that, although DESI identifies absorbers in large numbers, its lower spectral resolution results in line blending and possible mis-identifications, which limits its potential for comparisons with TNG. One could expect similar limitations with \textit{Euclid}'s NISP spectrometer that is sensitive to the Mg\,II doublet, but has a  resolving power of only $R \sim 450$ \citep{EuclidNISP}. 
\end{itemize}
More broadly, we demonstrate that simplified, grid-based ionization modeling can provide physically informative and computationally efficient predictions for metal-line observables in large cosmological simulations, enabling rapid exploration of ionization prescriptions and feedback scenarios. Future studies combining full spectral synthesis with higher-resolution simulations will enable a more robust interpretation of the strongest absorbers and provide deeper insight into the physical conditions of the CGM and IGM.

\section{Acknowledgements}
This research was supported by The Israel Science Foundation (grant No. 2617/25).
The authors thank Lucas Napolitano for useful discussions and comments on the manuscript. 
\texttt{CHIANTI} is a collaborative project involving George Mason University, the University of Michigan (USA), University of Cambridge (UK) and NASA Goddard Space Flight Center (USA).
This research used data obtained with the Dark Energy Spectroscopic Instrument (DESI). DESI construction and operations is managed by the Lawrence Berkeley National Laboratory. This material is based upon work supported by the U.S. Department of Energy, Office of Science, Office of High-Energy Physics, under Contract No. DE–AC02–05CH11231, and by the National Energy Research Scientific Computing Center, a DOE Office of Science User Facility under the same contract. Additional support for DESI was provided by the U.S. National Science Foundation (NSF), Division of Astronomical Sciences under Contract No. AST-0950945 to the NSF’s National Optical-Infrared Astronomy Research Laboratory; the Science and Technology Facilities Council of the United Kingdom; the Gordon and Betty Moore Foundation; the Heising-Simons Foundation; the French Alternative Energies and Atomic Energy Commission (CEA); the National Council of Humanities, Science and Technology of Mexico (CONAHCYT); the Ministry of Science and Innovation of Spain (MICINN), and by the DESI Member Institutions: www.desi.lbl.gov/collaborating-institutions. The DESI collaboration is honored to be permitted to conduct scientific research on I’oligam Du’ag (Kitt Peak), a mountain with particular significance to the Tohono O’odham Nation. Any opinions, findings, and conclusions or recommendations expressed in this material are those of the author(s) and do not necessarily reflect the views of the U.S. National Science Foundation, the U.S. Department of Energy, or any of the listed funding agencies.
\bibliography{references}

\begin{thebibliography}{}
\expandafter\ifx\csname natexlab\endcsname\relax\def\natexlab#1{#1}\fi
\providecommand{\url}[1]{\href{#1}{#1}}
\providecommand{\dodoi}[1]{doi:~\href{http://doi.org/#1}{\nolinkurl{#1}}}
\providecommand{\doeprint}[1]{\href{http://ascl.net/#1}{\nolinkurl{http://ascl.net/#1}}}
\providecommand{\doarXiv}[1]{\href{https://arxiv.org/abs/#1}{\nolinkurl{https://arxiv.org/abs/#1}}}

\bibitem[{Aghamousa {et~al.}(2016)}]{desicollaboration2016}
Aghamousa, A., {et~al.} 2016.
\newblock \doarXiv{1611.00036}

\bibitem[{Anand {et~al.}(2025)Anand, Aguilar, Ahlen, Bianchi, Brodzeller, Brooks, Canning, Claybaugh, Cuceu, de~la Macorra, Doel, Ferraro, Font-Ribera, Forero-Romero, Gaztañaga, Gontcho A~Gontcho, Gutierrez, Guy, Herrera-Alcantar, Ishak, Juneau, Kehoe, Kremin, Landriau, Le~Guillou, Levi, Manera, Meisner, Miquel, Moustakas, Muñoz-Gutiérrez, Napolitano, Pérez-Ràfols, Rossi, Sanchez, Schlegel, Schubnell, Sprayberry, Tarlé, Temple, Weaver, \& Zhou}]{Anand_2025}
Anand, A., Aguilar, J., Ahlen, S., {et~al.} 2025, The Astrophysical Journal, 990, 151, \dodoi{10.3847/1538-4357/adef3c}

\bibitem[{Artale {et~al.}(2021)Artale, Haider, Montero-Dorta, Vogelsberger, Martizzi, Torrey, Bird, Hernquist, \& Marinacci}]{Artale_2021}
Artale, M.~C., Haider, M., Montero-Dorta, A.~D., {et~al.} 2021, Monthly Notices of the Royal Astronomical Society, 510, 399, \dodoi{10.1093/mnras/stab3281}

\bibitem[{{Bahcall} \& {Salpeter}(1965)}]{bahcall/salpeter:1965}
{Bahcall}, J.~N., \& {Salpeter}, E.~E. 1965, \apj, 142, 1677, \dodoi{10.1086/148460}

\bibitem[{Barnes {et~al.}(2018)Barnes, Vogelsberger, Kannan, Marinacci, Weinberger, Springel, Torrey, Pillepich, Nelson, Pakmor, Naiman, Hernquist, \& McDonald}]{Barnes_2018}
Barnes, D.~J., Vogelsberger, M., Kannan, R., {et~al.} 2018, Monthly Notices of the Royal Astronomical Society, 481, 1809, \dodoi{10.1093/mnras/sty2078}

\bibitem[{{Bergeron} \& {Boiss{\'e}}(1991)}]{Bergeron_1991}
{Bergeron}, J., \& {Boiss{\'e}}, P. 1991, \aap, 243, 344

\bibitem[{{Bergeron} \& {Stasi{\'n}ska}(1986)}]{Bergron_Stasinska_1986}
{Bergeron}, J., \& {Stasi{\'n}ska}, G. 1986, \aap, 169, 1

\bibitem[{{Bouch{\'e}} {et~al.}(2012){Bouch{\'e}}, {Murphy}, {P{\'e}roux}, {Contini}, {Martin}, {Forster Schreiber}, {Genzel}, {Lutz}, {Gillessen}, {Tacconi}, {Davies}, \& {Eisenhauer}}]{Bouche_2012}
{Bouch{\'e}}, N., {Murphy}, M.~T., {P{\'e}roux}, C., {et~al.} 2012, \mnras, 419, 2, \dodoi{10.1111/j.1365-2966.2011.19500.x}

\bibitem[{{Chaussidon} {et~al.}(2023){Chaussidon}, {Y{\`e}che}, {Palanque-Delabrouille}, {Alexander}, {Yang}, {Ahlen}, {Bailey}, {Brooks}, {Cai}, {Chabanier}, {Davis}, {Dawson}, {de laMacorra}, {Dey}, {Dey}, {Eftekharzadeh}, {Eisenstein}, {Fanning}, {Font-Ribera}, {Gazta{\~n}aga}, {A Gontcho}, {Gonzalez-Morales}, {Guy}, {Herrera-Alcantar}, {Honscheid}, {Ishak}, {Jiang}, {Juneau}, {Kehoe}, {Kisner}, {Kov{\'a}cs}, {Kremin}, {Lan}, {Landriau}, {Le Guillou}, {Levi}, {Magneville}, {Martini}, {Meisner}, {Moustakas}, {Mu{\~n}oz-Guti{\'e}rrez}, {Myers}, {Newman}, {Nie}, {Percival}, {Poppett}, {Prada}, {Raichoor}, {Ravoux}, {Ross}, {Schlafly}, {Schlegel}, {Tan}, {Tarl{\'e}}, {Zhou}, {Zhou}, \& {Zou}}]{Chaussidon_2023}
{Chaussidon}, E., {Y{\`e}che}, C., {Palanque-Delabrouille}, N., {et~al.} 2023, \apj, 944, 107, \dodoi{10.3847/1538-4357/acb3c2}

\bibitem[{Chen {et~al.}(2010)Chen, Helsby, Gauthier, Shectman, Thompson, \& Tinker}]{Chen_2010}
Chen, H.-W., Helsby, J.~E., Gauthier, J.-R., {et~al.} 2010, The Astrophysical Journal, 714, 1521, \dodoi{10.1088/0004-637X/714/2/1521}

\bibitem[{Chen {et~al.}(2017)Chen, Simcoe, Torrey, Bañados, Cooksey, Cooper, Furesz, Matejek, Miller, Turner, Venemans, Decarli, Farina, Mazzucchelli, \& Walter}]{Chen_2017}
Chen, S.-F.~S., Simcoe, R.~A., Torrey, P., {et~al.} 2017, The Astrophysical Journal, 850, 188, \dodoi{10.3847/1538-4357/aa9707}

\bibitem[{Churchill {et~al.}(2025)Churchill, Abbas, Kacprzak, \& Nielsen}]{Churchill2025}
Churchill, C.~W., Abbas, A., Kacprzak, G.~G., \& Nielsen, N.~M. 2025, 13 Billion Years of MgII Absorber Evolution.
\newblock \doarXiv{2510.01430}

\bibitem[{Churchill {et~al.}(2020)Churchill, Evans, Stemock, Nielsen, Kacprzak, \& Murphy}]{Churchill_2020}
Churchill, C.~W., Evans, J.~L., Stemock, B., {et~al.} 2020, The Astrophysical Journal, 904, 28, \dodoi{10.3847/1538-4357/abbb34}

\bibitem[{Churchill {et~al.}(2000)Churchill, Mellon, Charlton, Jannuzi, Kirhakos, Steidel, \& Schneider}]{Churchill_2000}
Churchill, C.~W., Mellon, R.~R., Charlton, J.~C., {et~al.} 2000, The Astrophysical Journal Supplement Series, 130, 91–119, \dodoi{10.1086/317343}

\bibitem[{Churchill {et~al.}(2003)Churchill, Vogt, \& Charlton}]{Churchill_2003}
Churchill, C.~W., Vogt, S.~S., \& Charlton, J.~C. 2003, The Astronomical Journal, 125, 98, \dodoi{10.1086/345513}

\bibitem[{{Dawson} {et~al.}(2013){Dawson}, {Schlegel}, {Ahn}, {Anderson}, {Aubourg}, {Bailey}, {Barkhouser}, {Bautista}, {Beifiori}, {Berlind}, {Bhardwaj}, {Bizyaev}, {Blake}, {Blanton}, {Blomqvist}, {Bolton}, {Borde}, {Bovy}, {Brandt}, {Brewington}, {Brinkmann}, {Brown}, {Brownstein}, {Bundy}, {Busca}, {Carithers}, {Carnero}, {Carr}, {Chen}, {Comparat}, {Connolly}, {Cope}, {Croft}, {Cuesta}, {da Costa}, {Davenport}, {Delubac}, {de Putter}, {Dhital}, {Ealet}, {Ebelke}, {Eisenstein}, {Escoffier}, {Fan}, {Filiz Ak}, {Finley}, {Font-Ribera}, {G{\'e}nova-Santos}, {Gunn}, {Guo}, {Haggard}, {Hall}, {Hamilton}, {Harris}, {Harris}, {Ho}, {Hogg}, {Holder}, {Honscheid}, {Huehnerhoff}, {Jordan}, {Jordan}, {Kauffmann}, {Kazin}, {Kirkby}, {Klaene}, {Kneib}, {Le Goff}, {Lee}, {Long}, {Loomis}, {Lundgren}, {Lupton}, {Maia}, {Makler}, {Malanushenko}, {Malanushenko}, {Mandelbaum}, {Manera}, {Maraston}, {Margala}, {Masters}, {McBride}, {McDonald}, {McGreer}, {McMahon}, {Mena}, {Miralda-Escud{\'e}}, {Montero-Dorta},
  {Montesano}, {Muna}, {Myers}, {Naugle}, {Nichol}, {Noterdaeme}, {Nuza}, {Olmstead}, {Oravetz}, {Oravetz}, {Owen}, {Padmanabhan}, {Palanque-Delabrouille}, {Pan}, {Parejko}, {P{\^a}ris}, {Percival}, {P{\'e}rez-Fournon}, {P{\'e}rez-R{\`a}fols}, {Petitjean}, {Pfaffenberger}, {Pforr}, {Pieri}, {Prada}, {Price-Whelan}, {Raddick}, {Rebolo}, {Rich}, {Richards}, {Rockosi}, {Roe}, {Ross}, {Ross}, {Rossi}, {Rubi{\~n}o-Martin}, {Samushia}, {S{\'a}nchez}, {Sayres}, {Schmidt}, {Schneider}, {Sc{\'o}ccola}, {Seo}, {Shelden}, {Sheldon}, {Shen}, {Shu}, {Slosar}, {Smee}, {Snedden}, {Stauffer}, {Steele}, {Strauss}, {Streblyanska}, {Suzuki}, {Swanson}, {Tal}, {Tanaka}, {Thomas}, {Tinker}, {Tojeiro}, {Tremonti}, {Vargas Maga{\~n}a}, {Verde}, {Viel}, {Wake}, {Watson}, {Weaver}, {Weinberg}, {Weiner}, {West}, {White}, {Wood-Vasey}, {Yeche}, {Zehavi}, {Zhao}, \& {Zheng}}]{Dawson_2013}
{Dawson}, K.~S., {Schlegel}, D.~J., {Ahn}, C.~P., {et~al.} 2013, \aj, 145, 10, \dodoi{10.1088/0004-6256/145/1/10}

\bibitem[{{Del Zanna} {et~al.}(2015){Del Zanna}, {Dere}, {Young}, {Landi}, \& {Mason}}]{Chianti_2}
{Del Zanna}, G., {Dere}, K.~P., {Young}, P.~R., {Landi}, E., \& {Mason}, H.~E. 2015, \aap, 582, A56, \dodoi{10.1051/0004-6361/201526827}

\bibitem[{{Dere} {et~al.}(1997){Dere}, {Landi}, {Mason}, {Monsignori Fossi}, \& {Young}}]{Chianti_1}
{Dere}, K.~P., {Landi}, E., {Mason}, H.~E., {Monsignori Fossi}, B.~C., \& {Young}, P.~R. 1997, \aaps, 125, 149, \dodoi{10.1051/aas:1997368}

\bibitem[{{DESI Collaboration} {et~al.}(2025){DESI Collaboration}, {Abdul-Karim}, {Adame}, {et~al.}}]{DESI_DR1}
{DESI Collaboration}, {Abdul-Karim}, M., {Adame}, {et~al.} 2025, arXiv e-prints, arXiv:2503.14745, \dodoi{10.48550/arXiv.2503.14745}

\bibitem[{{Eisenstein} {et~al.}(2011){Eisenstein}, {Weinberg}, {Agol}, {Aihara}, {Allende Prieto}, {Anderson}, {Arns}, {Aubourg}, {Bailey}, {Balbinot}, {Barkhouser}, {Beers}, {Berlind}, {Bickerton}, {Bizyaev}, {Blanton}, {Bochanski}, {Bolton}, {Bosman}, {Bovy}, {Brandt}, {Breslauer}, {Brewington}, {Brinkmann}, {Brown}, {Brownstein}, {Burger}, {Busca}, {Campbell}, {Cargile}, {Carithers}, {Carlberg}, {Carr}, {Chang}, {Chen}, {Chiappini}, {Comparat}, {Connolly}, {Cortes}, {Croft}, {Cunha}, {da Costa}, {Davenport}, {Dawson}, {De Lee}, {Porto de Mello}, {de Simoni}, {Dean}, {Dhital}, {Ealet}, {Ebelke}, {Edmondson}, {Eiting}, {Escoffier}, {Esposito}, {Evans}, {Fan}, {Femen{\'\i}a Castell{\'a}}, {Dutra Ferreira}, {Fitzgerald}, {Fleming}, {Font-Ribera}, {Ford}, {Frinchaboy}, {Garc{\'\i}a P{\'e}rez}, {Gaudi}, {Ge}, {Ghezzi}, {Gillespie}, {Gilmore}, {Girardi}, {Gott}, {Gould}, {Grebel}, {Gunn}, {Hamilton}, {Harding}, {Harris}, {Hawley}, {Hearty}, {Hennawi}, {Gonz{\'a}lez Hern{\'a}ndez}, {Ho}, {Hogg}, {Holtzman},
  {Honscheid}, {Inada}, {Ivans}, {Jiang}, {Jiang}, {Johnson}, {Jordan}, {Jordan}, {Kauffmann}, {Kazin}, {Kirkby}, {Klaene}, {Knapp}, {Kneib}, {Kochanek}, {Koesterke}, {Kollmeier}, {Kron}, {Lampeitl}, {Lang}, {Lawler}, {Le Goff}, {Lee}, {Lee}, {Leisenring}, {Lin}, {Liu}, {Long}, {Loomis}, {Lucatello}, {Lundgren}, {Lupton}, {Ma}, {Ma}, {MacDonald}, {Mack}, {Mahadevan}, {Maia}, {Majewski}, {Makler}, {Malanushenko}, {Malanushenko}, {Mandelbaum}, {Maraston}, {Margala}, {Maseman}, {Masters}, {McBride}, {McDonald}, {McGreer}, {McMahon}, {Mena Requejo}, {M{\'e}nard}, {Miralda-Escud{\'e}}, {Morrison}, {Mullally}, {Muna}, {Murayama}, {Myers}, {Naugle}, {Neto}, {Nguyen}, {Nichol}, {Nidever}, {O'Connell}, {Ogando}, {Olmstead}, {Oravetz}, {Padmanabhan}, {Paegert}, {Palanque-Delabrouille}, {Pan}, {Pandey}, {Parejko}, {P{\^a}ris}, {Pellegrini}, {Pepper}, {Percival}, {Petitjean}, {Pfaffenberger}, {Pforr}, {Phleps}, {Pichon}, {Pieri}, {Prada}, {Price-Whelan}, {Raddick}, {Ramos}, {Reid}, {Reyle}, {Rich}, {Richards}, {Rieke},
  {Rieke}, {Rix}, {Robin}, {Rocha-Pinto}, {Rockosi}, {Roe}, {Rollinde}, {Ross}, {Ross}, {Rossetto}, {S{\'a}nchez}, {Santiago}, {Sayres}, {Schiavon}, {Schlegel}, {Schlesinger}, {Schmidt}, {Schneider}, {Sellgren}, {Shelden}, {Sheldon}, \& {Shetrone}}]{Eisenstein_2011}
{Eisenstein}, D.~J., {Weinberg}, D.~H., {Agol}, E., {et~al.} 2011, \aj, 142, 72, \dodoi{10.1088/0004-6256/142/3/72}

\bibitem[{{Euclid Collaboration} {et~al.}(2025){Euclid Collaboration}, {Jahnke, K.}, {Gillard, W.}, {Schirmer, M.}, {Ealet, A.}, {Maciaszek, T.}, {Prieto, E.}, {Barbier, R.}, {Bonoli, C.}, {Corcione, L.}, {Dusini, S.}, {Grupp, F.}, {Hormuth, F.}, {Ligori, S.}, {Martin, L.}, {Morgante, G.}, {Padilla, C.}, {Toledo-Moreo, R.}, {Trifoglio, M.}, {Valenziano, L.}, {Bender, R.}, {Castander, F. J.}, {Garilli, B.}, {Lilje, P. B.}, {Rix, H.-W.}, {Andersen, M. I.}, {Auricchio, N.}, {Balestra, A.}, {Barriere, J.-C.}, {Battaglia, P.}, {Berthe, M.}, {Bodendorf, C.}, {Boenke, T.}, {Bon, W.}, {Bonnefoi, A.}, {Caillat, A.}, {Capobianco, V.}, {Carle, M.}, {Casas, R.}, {Cho, H.}, {Costille, A.}, {Ducret, F.}, {Ferriol, S.}, {Franceschi, E.}, {Gimenez, J.-L.}, {Holmes, W.}, {Hornstrup, A.}, {Jhabvala, M.}, {Kohley, R.}, {Kubik, B.}, {Laureijs, R.}, {Le Mignant, D.}, {Lloro, I.}, {Medinaceli, E.}, {Mellier, Y.}, {Polenta, G.}, {Racca, G. D.}, {Renzi, A.}, {Salvignol, J.-C.}, {Secroun, A.}, {Seidel, G.}, {Seiffert, M.},
  {Sirignano, C.}, {Sirri, G.}, {Strada, P.}, {Smadja, G.}, {Stanco, L.}, {Wachter, S.}, {Anselmi, S.}, {Borsato, E.}, {Caillat, L.}, {Cogato, F.}, {Colodro-Conde, C.}, {Crouzet, P.-E.}, {Conforti, V.}, {D\'{}Alessandro, M.}, {Copin, Y.}, {Cuillandre, J.-C.}, {Davies, J. E.}, {Davini, S.}, {Derosa, A.}, {Diaz, J. J.}, {Di Domizio, S.}, {Di Ferdinando, D.}, {Farinelli, R.}, {Ferrari, A. G.}, {Fornari, F.}, {Gabarra, L.}, {Garcia, R.}, {Gutierrez, C. M.}, {Giacomini, F.}, {Lagier, P.}, {Gianotti, F.}, {Krause, O.}, {Madrid, F.}, {Laudisio, F.}, {Macias-Perez, J.}, {Naletto, G.}, {Niclas, M.}, {Marpaud, J.}, {Mauri, N.}, {da Silva, R.}, {Passalacqua, F.}, {Paterson, K.}, {Patrizii, L.}, {Risso, I.}, {Solheim, B. G. B.}, {Scodeggio, M.}, {Stassi, P.}, {Steinwagner, J.}, {Tenti, M.}, {Testera, G.}, {Travaglini, R.}, {Tosi, S.}, {Troja, A.}, {Tubio, O.}, {Valieri, C.}, {Vescovi, C.}, {Ventura, S.}, {Aghanim, N.}, {Altieri, B.}, {Amara, A.}, {Amiaux, J.}, {Andreon, S.}, {Appleton, P. N.}, {Aussel, H.}, {Baccigalupi,
  C.}, {Baldi, M.}, {Bardelli, S.}, {Basset, A.}, {Bonchi, A.}, {Bonino, D.}, {Branchini, E.}, {Brescia, M.}, {Brinchmann, J.}, {Camera, S.}, {Carbone, C.}, {Cardone, V. F.}, {Carretero, J.}, {Casas, S.}, {Castellano, M.}, {Castignani, G.}, {Cavuoti, S.}, {Chabaud, P.-Y.}, {Cimatti, A.}, {Congedo, G.}, {Conselice, C. J.}, {Conversi, L.}, {Courbin, F.}, {Courtois, H. M.}, {Crocce, M.}, {Cropper, M.}, {Cuby, J.-G.}, {Da Silva, A.}, {Degaudenzi, H.}, {De Lucia, G.}, {Di Giorgio, A. M.}, {Dinis, J.}, {Douspis, M.}, {Dubath, F.}, {Duncan, C. A. J.}, {Dupac, X.}, {Fabricius, M.}, {Farina, M.}, {Farrens, S.}, {Faustini, F.}, {Fosalba, P.}, {Fotopoulou, S.}, {Fourmanoit, N.}, {Frailis, M.}, {Franzetti, P.}, {Galeotta, S.}, {George, K.}, {Gillis, B.}, {Giocoli, C.}, {G\'omez-Alvarez, P.}, {Granett, B. R.}, {Grazian, A.}, {Guzzo, L.}, {Hailey, M.}, {Haugan, S. V. H.}, {Hoar, J.}, {Hoekstra, H.}, {Hook, I.}, {Hudelot, P.}, {Ili\'{}c, S.}, {Joachimi, B.}, {Keih\"anen, E.}, {Kermiche, S.}, {Kiessling, A.}, {Kilbinger,
  M.}, {Kitching, T.}, {K\"ummel, M.}, {Kunz, M.}, {Kurki-Suonio, H.}, {Lahav, O.}, {Liebing, P.}, {Lindholm, V.}, {Lorenzo Alvarez, J.}, {Mainetti, G.}, {Maino, D.}, {Maiorano, E.}, {Mansutti, O.}, {Marcin, S.}, {Marggraf, O.}, {Markovic, K.}, {Martignac, J.}, {Martinelli, M.}, {Martinet, N.}, {Marulli, F.}, {Massey, R.}, {Masters, D. C.}, {Maurogordato, S.}, {McCracken, H. J.}, {Mei, S.}, {Melchior, M.}, {Meneghetti, M.}, {Merlin, E.}, {Meylan, G.}, {Mohr, J. J.}, {Moresco, M.}, {Morris, P. W.}, {Moscardini, L.}, {Nakajima, R.}, {Neissner, C.}, {Nichol, R. C.}, {Niemi, S.-M.}, {Nutma, T.}, {Paech, K.}, {Paltani, S.}, {Pasian, F.}, {Peacock, J. A.}, {Pedersen, K.}, {Percival, W. J.}, {Pettorino, V.}, {Pires, S.}, {Poncet, M.}, {Popa, L. A.}, {Pozzetti, L.}, {Raison, F.}, {Rebolo, R.}, {Refregier, A.}, {Rhodes, J.}, {Riccio, G.}, {Romelli, E.}, {Roncarelli, M.}, {Rosset, C.}, {Rossetti, E.}, {Rottgering, H. J. A.}, {Rusholme, B.}, {Saglia, R.}, {Sakr, Z.}, {S\'anchez, A. G.}, {Sapone, D.}, {Sauvage, M.},
  {Scaramella, R.}, {Schewtschenko, J. A.}, {Schneider, P.}, {Schrabback, T.}, {Sefusatti, E.}, {Serrano, S.}, {Tallada-Cresp\'{\i}, P.}, {Tavagnacco, D.}, {Taylor, A. N.}, {Teplitz, H. I.}, {Tereno, I.}, {Torradeflot, F.}, {Tutusaus, I.}, {Vassallo, T.}, {Verdoes Kleijn, G.}, {Veropalumbo, A.}, {Vibert, D.}, {Wang, Y.}, {Weller, J.}, {Zacchei, A.}, {Zamorani, G.}, {Zerbi, F. M.}, {Zoubian, J.}, {Zucca, E.}, {Biviano, A.}, {Bolzonella, M.}, {Boucaud, A.}, {Bozzo, E.}, {Burigana, C.}, {Calabrese, M.}, {Casenove, P.}, {Escartin Vigo, J. A.}, {Fabbian, G.}, {Finelli, F.}, {Gracia-Carpio, J.}, {Liu, C.}, {Pezzotta, A.}, {P\"ontinen, M.}, {Porciani, C.}, {Scottez, V.}, {Viel, M.}, {Wiesmann, M.}, {Akrami, Y.}, {Allevato, V.}, {Aubourg, E.}, {Ballardini, M.}, {Bertacca, D.}, {Bethermin, M.}, {Blanchard, A.}, {Blot, L.}, {Borgani, S.}, {Borlaff, A. S.}, {Bruton, S.}, {Cabanac, R.}, {Calabro, A.}, {Calderone, G.}, {Canas-Herrera, G.}, {Cappi, A.}, {Carvalho, C. S.}, {Castro, T.}, {Chambers, K. C.}, {Charles, Y.},
  {Chary, R.}, {Colbert, J.}, {Contarini, S.}, {Contini, T.}, {Cooray, A. R.}, {Costanzi, M.}, {Cucciati, O.}, {De Caro, B.}, {de la Torre, S.}, {Desprez, G.}, {D\'{\i}az-S\'anchez, A.}, {Dole, H.}, {Escoffier, S.}, {Ferreira, P. G.}, {Ferrero, I.}, {Finoguenov, A.}, {Fontana, A.}, {Ganga, K.}, {Garc\'{\i}a-Bellido, J.}, {Gautard, V.}, {Gaztanaga, E.}, {Gozaliasl, G.}, {Gregorio, A.}, {Hall, A.}, {Hartley, W. G.}, {Hemmati, S.}, {Hildebrandt, H.}, {Hjorth, J.}, {Hosseini, S.}, {Huertas-Company, M.}, {Ilbert, O.}, {Jacobson, J.}, {Jimenez Mu\~noz, A.}, {Joudaki, S.}, {Kajava, J. J. E.}, {Kansal, V.}, {Karagiannis, D.}, {Kirkpatrick, C. C.}, {Lacasa, F.}, {Le Brun, V.}, {Le Graet, J.}, {Legrand, L.}, {Libet, G.}, {Liu, S. J.}, {Loureiro, A.}, {Magliocchetti, M.}, {Mancini, C.}, {Mannucci, F.}, {Maoli, R.}, {Martins, C. J. A. P.}, {Matthew, S.}, {Maurin, L.}, {McPartland, C. J. R.}, {Metcalf, R. B.}, {Migliaccio, M.}, {Miluzio, M.}, {Monaco, P.}, {Moretti, C.}, {Nadathur, S.}, {Nicastro, L.}, {Walton, Nicholas
  A.}, {Odier, J.}, {Oguri, M.}, {Popa, V.}, {Potter, D.}, {Pourtsidou, A.}, {Rocci, P.-F.}, {Rollins, R. P.}, {Sahl\'en, M.}, {Scarlata, C.}, {Schaye, J.}, {Schneider, A.}, {Schultheis, M.}, {Sereno, M.}, {Shankar, F.}, {Shulevski, A.}, {Sikkema, G.}, {Silvestri, A.}, {Simon, P.}, {Spurio Mancini, A.}, {Stadel, J.}, {Stanford, S. A.}, {Tanidis, K.}, {Tao, C.}, {Tessore, N.}, {Teyssier, R.}, {Toft, S.}, {Tucci, M.}, {Valiviita, J.}, {Vergani, D.}, {Vernizzi, F.}, {Verza, G.}, {Vielzeuf, P.}, {Weaver, J. R.}, {Zalesky, L.}, {Zinchenko, I. A.}, {Archidiacono, M.}, {Atrio-Barandela, F.}, {Bennett, C. L.}, {Bouvard, T.}, {Caro, F.}, {Conseil, S.}, {Dimauro, P.}, {Duc, P.-A.}, {Fang, Y.}, {Ferguson, A. M. N.}, {Gasparetto, T.}, {Kovaci\'{}c, I.}, {Kruk, S.}, {Le Brun, A. M. C.}, {Liaudat, T. I.}, {Montoro, A.}, {Mora, A.}, {Murray, C.}, {Pagano, L.}, {Paoletti, D.}, {Radovich, M.}, {Sarpa, E.}, {Tommasi, E.}, {Viitanen, A.}, {Lesgourgues, J.}, {Levi, M. E.}, {Mart\'{\i}n-Fleitas, J.}, \& {Oppizzi,
  F.}}]{EuclidNISP}
{Euclid Collaboration}, {Jahnke, K.}, {Gillard, W.}, {et~al.} 2025, A\&A, 697, A3, \dodoi{10.1051/0004-6361/202450786}

\bibitem[{{Faucher-Gigu{\`e}re} {et~al.}(2009){Faucher-Gigu{\`e}re}, {Lidz}, {Zaldarriaga}, \& {Hernquist}}]{Faucher_Giguere_2009}
{Faucher-Gigu{\`e}re}, C.-A., {Lidz}, A., {Zaldarriaga}, M., \& {Hernquist}, L. 2009, \apj, 703, 1416, \dodoi{10.1088/0004-637X/703/2/1416}

\bibitem[{{Gauthier} {et~al.}(2014){Gauthier}, {Chen}, {Cooksey}, {Simcoe}, {Seyffert}, \& {O'Meara}}]{Gauthier_2014}
{Gauthier}, J.-R., {Chen}, H.-W., {Cooksey}, K.~L., {et~al.} 2014, \mnras, 439, 342, \dodoi{10.1093/mnras/stt2443}

\bibitem[{{Grauer} \& {Behar}(2023)}]{Grauer2023}
{Grauer}, M., \& {Behar}, E. 2023, \apj, 953, 158, \dodoi{10.3847/1538-4357/ace1e8}

\bibitem[{{Gunn} \& {Peterson}(1965)}]{gunn/peterson:1965}
{Gunn}, J.~E., \& {Peterson}, B.~A. 1965, \apj, 142, 1633, \dodoi{10.1086/148444}

\bibitem[{Hafen {et~al.}(2019)Hafen, Faucher-Giguère, Anglés-Alcázar, Stern, Kereš, Hummels, Esmerian, Garrison-Kimmel, El-Badry, Wetzel, Chan, Hopkins, \& Murray}]{Hafen_2019}
Hafen, Z., Faucher-Giguère, C.-A., Anglés-Alcázar, D., {et~al.} 2019, Monthly Notices of the Royal Astronomical Society, 488, 1248, \dodoi{10.1093/mnras/stz1773}

\bibitem[{Hummels {et~al.}(2019)Hummels, Smith, Hopkins, O’Shea, Silvia, Werk, Lehner, Wise, Collins, \& Butsky}]{Hummels_2019}
Hummels, C.~B., Smith, B.~D., Hopkins, P.~F., {et~al.} 2019, The Astrophysical Journal, 882, 156, \dodoi{10.3847/1538-4357/ab378f}

\bibitem[{Jannuzi {et~al.}(1998)Jannuzi, Bahcall, Bergeron, Boksenberg, Hartig, Kirhakos, Sargent, Savage, Schneider, Turnshek, Weymann, \& Wolfe}]{Jannuzi_1998}
Jannuzi, B.~T., Bahcall, J.~N., Bergeron, J., {et~al.} 1998, The Astrophysical Journal Supplement Series, 118, 1–125, \dodoi{10.1086/313130}

\bibitem[{Kacprzak {et~al.}(2008)Kacprzak, Churchill, Steidel, \& Murphy}]{Kacprzak_2008}
Kacprzak, G.~G., Churchill, C.~W., Steidel, C.~C., \& Murphy, M.~T. 2008, The Astronomical Journal, 135, 922, \dodoi{10.1088/0004-6256/135/3/922}

\bibitem[{Kakoly {et~al.}(2025)Kakoly, Stern, Faucher-Giguère, Fielding, Goldner, Sun, \& Hummels}]{Kakoly_2025}
Kakoly, A., Stern, J., Faucher-Giguère, C.-A., {et~al.} 2025, Monthly Notices of the Royal Astronomical Society, 543, 3345, \dodoi{10.1093/mnras/staf1516}

\bibitem[{Khaire {et~al.}(2023)Khaire, Hu, Hennawi, Walther, \& Davies}]{Khaire_2023}
Khaire, V., Hu, T., Hennawi, J.~F., Walther, M., \& Davies, F. 2023, Monthly Notices of the Royal Astronomical Society, 527, 4545, \dodoi{10.1093/mnras/stad3374}

\bibitem[{{Kim} {et~al.}(2013){Kim}, {Partl}, {Carswell}, \& {M{\"u}ller}}]{Kim_2013}
{Kim}, T.-S., {Partl}, A.~M., {Carswell}, R.~F., \& {M{\"u}ller}, V. 2013, \aap, 552, A77, \dodoi{10.1051/0004-6361/201220042}

\bibitem[{Lan \& Fukugita(2017)}]{Lan_2017}
Lan, T.-W., \& Fukugita, M. 2017, The Astrophysical Journal, 850, 156, \dodoi{10.3847/1538-4357/aa93eb}

\bibitem[{{Lan} {et~al.}(2014){Lan}, {M{\'e}nard}, \& {Zhu}}]{Lan_2014}
{Lan}, T.-W., {M{\'e}nard}, B., \& {Zhu}, G. 2014, \apj, 795, 31, \dodoi{10.1088/0004-637X/795/1/31}

\bibitem[{Lan {et~al.}(2025)Lan, Prochaska, Aguilar, Ahlen, Anand, Bianchi, Brooks, Castander, Claybaugh, de~la Macorra, Doel, Ferraro, Font-Ribera, Forero-Romero, Gaztañaga, Gutierrez, Joyce, Juneau, Kehoe, Kisner, Kremin, Landriau, Guillou, Manera, Meisner, Miquel, Moustakas, Nadathur, Percival, Prada, Pérez-Ràfols, Rossi, Sanchez, Schlegel, Schubnell, Seo, Silber, Sprayberry, Tarlé, Weaver, Zhou, \& Zou}]{Lan2025}
Lan, T.-W., Prochaska, J.~X., Aguilar, J., {et~al.} 2025, The Multi-Phase Circumgalactic Medium of DESI Emission-Line Galaxies at z~1.5.
\newblock \doarXiv{2511.03195}

\bibitem[{Lucchini {et~al.}(2026)Lucchini, Abramson, Hummels, Conroy, Hernquist, \& Smith}]{Lucchini2026}
Lucchini, S., Abramson, C., Hummels, C., {et~al.} 2026, ENhanced Galactic Atmospheres With Arepo: Resolving the CGM at 200 pc with the ENGAWA Simulations.
\newblock \doarXiv{2603.05584}

\bibitem[{{Lyke} {et~al.}(2020){Lyke}, {Higley}, {McLane}, {Schurhammer}, {Myers}, {Ross}, {Dawson}, {Chabanier}, {Martini}, {Busca}, {Mas des Bourboux}, {Salvato}, {Streblyanska}, {Zarrouk}, {Burtin}, {Anderson}, {Bautista}, {Bizyaev}, {Brandt}, {Brinkmann}, {Brownstein}, {Comparat}, {Green}, {de la Macorra}, {Mu{\~n}oz Guti{\'e}rrez}, {Hou}, {Newman}, {Palanque-Delabrouille}, {P{\^a}ris}, {Percival}, {Petitjean}, {Rich}, {Rossi}, {Schneider}, {Smith}, {Vivek}, \& {Weaver}}]{Lyke_2020}
{Lyke}, B.~W., {Higley}, A.~N., {McLane}, J.~N., {et~al.} 2020, \apjs, 250, 8, \dodoi{10.3847/1538-4365/aba623}

\bibitem[{Marinacci {et~al.}(2018)Marinacci, Vogelsberger, Pakmor, Torrey, Springel, Hernquist, Nelson, Weinberger, Pillepich, Naiman, \& Genel}]{TNG_2}
Marinacci, F., Vogelsberger, M., Pakmor, R., {et~al.} 2018, Monthly Notices of the Royal Astronomical Society, \dodoi{10.1093/mnras/sty2206}

\bibitem[{Matejek \& Simcoe(2012)}]{Matejek_2012}
Matejek, M.~S., \& Simcoe, R.~A. 2012, The Astrophysical Journal, 761, 112, \dodoi{10.1088/0004-637X/761/2/112}

\bibitem[{Mathes {et~al.}(2017)Mathes, Churchill, \& Murphy}]{Mathes_2017}
Mathes, N.~L., Churchill, C.~W., \& Murphy, M.~T. 2017, The Vulture Survey I: Analyzing the Evolution of Mg\,II Absorbers.
\newblock \doarXiv{1701.05624}

\bibitem[{Meiksin(2009)}]{Meiksin_2009_IGM}
Meiksin, A.~A. 2009, Rev. Mod. Phys., 81, 1405, \dodoi{10.1103/RevModPhys.81.1405}

\bibitem[{{Monaghan}(1992)}]{Monaghan_1992}
{Monaghan}, J.~J. 1992, \araa, 30, 543, \dodoi{10.1146/annurev.aa.30.090192.002551}

\bibitem[{Naiman {et~al.}(2018)Naiman, Pillepich, Springel, Ramirez-Ruiz, Torrey, Vogelsberger, Pakmor, Nelson, Marinacci, Hernquist, Weinberger, \& Genel}]{TNG_3}
Naiman, J.~P., Pillepich, A., Springel, V., {et~al.} 2018, Monthly Notices of the Royal Astronomical Society, 477, 1206–1224, \dodoi{10.1093/mnras/sty618}

\bibitem[{{Napolitano} {et~al.}(2023){Napolitano}, {Pandey}, {Myers}, {Lan}, {Anand}, {Aguilar}, {Ahlen}, {Alexander}, {Brooks}, {Canning}, {Circosta}, {De La Macorra}, {Doel}, {Eftekharzadeh}, {Fawcett}, {Font-Ribera}, {Garcia-Bellido}, {Gontcho A Gontcho}, {Le Guillou}, {Guy}, {Honscheid}, {Juneau}, {Kisner}, {Landriau}, {Meisner}, {Miquel}, {Moustakas}, {Percival}, {Prochaska}, {Schubnell}, {Tarl{\'e}}, {Weaver}, {Weiner}, {Zhou}, {Zou}, \& {Zou}}]{Napolitano_2023}
{Napolitano}, L., {Pandey}, A., {Myers}, A.~D., {et~al.} 2023, \aj, 166, 99, \dodoi{10.3847/1538-3881/ace62c}

\bibitem[{{Napolitano} {et~al.}(2025){Napolitano}, {Myers}, {Fawcett}, {Aguilar}, {Ahlen}, {Bianchi}, {Brooks}, {Claybaugh}, {Cole}, {de la Macorra}, {Dey}, {Font-Ribera}, {Forero-Romero}, {Gazta{\~n}aga}, {Gontcho A Gontcho}, {Gutierrez}, {Honscheid}, {Juneau}, {Lambert}, {Landriau}, {Le Guillou}, {Meisner}, {Miquel}, {Moustakas}, {Newman}, {Prada}, {P{\'e}rez-R{\`a}fols}, {Rossi}, {Sanchez}, {Schlegel}, {Schubnell}, {Sprayberry}, {Tarl{\'e}}, {Weaver}, \& {Zou}}]{Napolitano_2025}
{Napolitano}, L., {Myers}, A.~D., {Fawcett}, V.~A., {et~al.} 2025, \aj, 170, 16, \dodoi{10.3847/1538-3881/adc389}

\bibitem[{Narayanan {et~al.}(2005)Narayanan, Charlton, Masiero, \& Lynch}]{Narayanan_2005}
Narayanan, A., Charlton, J.~C., Masiero, J.~R., \& Lynch, R. 2005, The Astrophysical Journal, 632, 92, \dodoi{10.1086/432750}

\bibitem[{Narayanan {et~al.}(2007)Narayanan, Misawa, Charlton, \& Kim}]{Narayanan_2007}
Narayanan, A., Misawa, T., Charlton, J.~C., \& Kim, T. 2007, The Astrophysical Journal, 660, 1093–1105, \dodoi{10.1086/512852}

\bibitem[{Nelson {et~al.}(2021{\natexlab{a}})Nelson, Byrohl, Peroux, Rubin, \& Burchett}]{Nelson_2021}
Nelson, D., Byrohl, C., Peroux, C., Rubin, K. H.~R., \& Burchett, J.~N. 2021{\natexlab{a}}, Monthly Notices of the Royal Astronomical Society, 507, 4445–4463, \dodoi{10.1093/mnras/stab2177}

\bibitem[{Nelson {et~al.}(2017)Nelson, Pillepich, Springel, Weinberger, Hernquist, Pakmor, Genel, Torrey, Vogelsberger, Kauffmann, Marinacci, \& Naiman}]{TNG_1}
Nelson, D., Pillepich, A., Springel, V., {et~al.} 2017, Monthly Notices of the Royal Astronomical Society, 475, 624–647, \dodoi{10.1093/mnras/stx3040}

\bibitem[{Nelson {et~al.}(2018)Nelson, Kauffmann, Pillepich, Genel, Springel, Pakmor, Hernquist, Weinberger, Torrey, Vogelsberger, \& Marinacci}]{Nelson_2018}
Nelson, D., Kauffmann, G., Pillepich, A., {et~al.} 2018, Monthly Notices of the Royal Astronomical Society, 477, 450–479, \dodoi{10.1093/mnras/sty656}

\bibitem[{Nelson {et~al.}(2019)Nelson, Pillepich, Springel, Pakmor, Weinberger, Genel, Torrey, Vogelsberger, Marinacci, \& Hernquist}]{TNG50a}
Nelson, D., Pillepich, A., Springel, V., {et~al.} 2019, Monthly Notices of the Royal Astronomical Society, 490, 3234, \dodoi{10.1093/mnras/stz2306}

\bibitem[{Nelson {et~al.}(2020)Nelson, Sharma, Pillepich, Springel, Pakmor, Weinberger, Vogelsberger, Marinacci, \& Hernquist}]{Nelson_2020}
Nelson, D., Sharma, P., Pillepich, A., {et~al.} 2020, Monthly Notices of the Royal Astronomical Society, 498, 2391, \dodoi{10.1093/mnras/staa2419}

\bibitem[{Nelson {et~al.}(2021{\natexlab{b}})Nelson, Springel, Pillepich, Rodriguez-Gomez, Torrey, Genel, Vogelsberger, Pakmor, Marinacci, Weinberger, Kelley, Lovell, Diemer, \& Hernquist}]{TNG_DR}
Nelson, D., Springel, V., Pillepich, A., {et~al.} 2021{\natexlab{b}}, The IllustrisTNG Simulations: Public Data Release.
\newblock \doarXiv{1812.05609}

\bibitem[{Nelson {et~al.}(2025)Nelson, Peroux, Richter, Pieri, Lopez, Bordoloi, Zou, Burchett, Davies, Ramesh, Smith, Borthakur, \& Churchill}]{nelson2025_salsa}
Nelson, D., Peroux, C., Richter, P., {et~al.} 2025, The Synthetic Absorption Line Spectral Almanac (SALSA).
\newblock \doarXiv{2510.19904}

\bibitem[{Nestor {et~al.}(2005)Nestor, Turnshek, \& Rao}]{Nestor_2005}
Nestor, D.~B., Turnshek, D.~A., \& Rao, S.~M. 2005, The Astrophysical Journal, 628, 637, \dodoi{10.1086/427547}

\bibitem[{{Noterdaeme} {et~al.}(2012){Noterdaeme}, {Petitjean}, {Carithers}, {P{\^a}ris}, {Font-Ribera}, {Bailey}, {Aubourg}, {Bizyaev}, {Ebelke}, {Finley}, {Ge}, {Malanushenko}, {Malanushenko}, {Miralda-Escud{\'e}}, {Myers}, {Oravetz}, {Pan}, {Pieri}, {Ross}, {Schneider}, {Simmons}, \& {York}}]{Noterdaeme_2012}
{Noterdaeme}, P., {Petitjean}, P., {Carithers}, W.~C., {et~al.} 2012, \aap, 547, L1, \dodoi{10.1051/0004-6361/201220259}

\bibitem[{{Oppenheimer} {et~al.}(2018){Oppenheimer}, {Segers}, {Schaye}, {Richings}, \& {Crain}}]{Oppenheimer_2018}
{Oppenheimer}, B.~D., {Segers}, M., {Schaye}, J., {Richings}, A.~J., \& {Crain}, R.~A. 2018, \mnras, 474, 4740, \dodoi{10.1093/mnras/stx2967}

\bibitem[{Pieri {et~al.}(2014)Pieri, Mortonson, Frank, Crighton, Weinberg, Lee, Noterdaeme, Bailey, Busca, Ge, Kirkby, Lundgren, Mathur, Pâris, Palanque-Delabrouille, Petitjean, Rich, Ross, Schneider, \& York}]{Pieri_2014}
Pieri, M.~M., Mortonson, M.~J., Frank, S., {et~al.} 2014, Monthly Notices of the Royal Astronomical Society, 441, 1718, \dodoi{10.1093/mnras/stu577}

\bibitem[{Pillepich {et~al.}(2017{\natexlab{a}})Pillepich, Nelson, Hernquist, Springel, Pakmor, Torrey, Weinberger, Genel, Naiman, Marinacci, \& Vogelsberger}]{TNG_5}
Pillepich, A., Nelson, D., Hernquist, L., {et~al.} 2017{\natexlab{a}}, Monthly Notices of the Royal Astronomical Society, 475, 648–675, \dodoi{10.1093/mnras/stx3112}

\bibitem[{Pillepich {et~al.}(2017{\natexlab{b}})Pillepich, Springel, Nelson, Genel, Naiman, Pakmor, Hernquist, Torrey, Vogelsberger, Weinberger, \& Marinacci}]{TNG_7}
Pillepich, A., Springel, V., Nelson, D., {et~al.} 2017{\natexlab{b}}, Monthly Notices of the Royal Astronomical Society, 473, 4077–4106, \dodoi{10.1093/mnras/stx2656}

\bibitem[{Pillepich {et~al.}(2019)Pillepich, Nelson, Springel, Pakmor, Torrey, Weinberger, Vogelsberger, Marinacci, Genel, van der Wel, \& Hernquist}]{TNG50b}
Pillepich, A., Nelson, D., Springel, V., {et~al.} 2019, Monthly Notices of the Royal Astronomical Society, 490, 3196, \dodoi{10.1093/mnras/stz2338}

\bibitem[{{Press} {et~al.}(1993){Press}, {Rybicki}, \& {Schneider}}]{press/etal:1993}
{Press}, W.~H., {Rybicki}, G.~B., \& {Schneider}, D.~P. 1993, \apj, 414, 64, \dodoi{10.1086/173057}

\bibitem[{Prochaska {et~al.}(2008)Prochaska, Hennawi, \& Herbert-Fort}]{Prochaska_2008}
Prochaska, J.~X., Hennawi, J.~F., \& Herbert-Fort, S. 2008, The Astrophysical Journal, 675, 1002, \dodoi{10.1086/526508}

\bibitem[{Péroux {et~al.}(2020)Péroux, Nelson, van de Voort, Pillepich, Marinacci, Vogelsberger, \& Hernquist}]{Peroux_2020}
Péroux, C., Nelson, D., van de Voort, F., {et~al.} 2020, Monthly Notices of the Royal Astronomical Society, 499, 2462, \dodoi{10.1093/mnras/staa2888}

\bibitem[{Quider {et~al.}(2011)Quider, Nestor, Turnshek, Rao, Monier, Weyant, \& Busche}]{Quider_2011}
Quider, A.~M., Nestor, D.~B., Turnshek, D.~A., {et~al.} 2011, The Astronomical Journal, 141, 137, \dodoi{10.1088/0004-6256/141/4/137}

\bibitem[{Rauch(1998)}]{Rauch_1998}
Rauch, M. 1998, Annual Review of Astronomy and Astrophysics, 36, 267–316, \dodoi{10.1146/annurev.astro.36.1.267}

\bibitem[{Rigby {et~al.}(2002)Rigby, Charlton, \& Churchill}]{Rigby_2002}
Rigby, J.~R., Charlton, J.~C., \& Churchill, C.~W. 2002, The Astrophysical Journal, 565, 743, \dodoi{10.1086/324723}

\bibitem[{Seyffert {et~al.}(2013)Seyffert, Cooksey, Simcoe, O'Meara, Kao, \& Prochaska}]{Seyffert_2013}
Seyffert, E.~N., Cooksey, K.~L., Simcoe, R.~A., {et~al.} 2013, The Astrophysical Journal, 779, 161, \dodoi{10.1088/0004-637X/779/2/161}

\bibitem[{Springel \& Hernquist(2003)}]{Springel_2003}
Springel, V., \& Hernquist, L. 2003, Monthly Notices of the Royal Astronomical Society, 339, 289–311, \dodoi{10.1046/j.1365-8711.2003.06206.x}

\bibitem[{Springel {et~al.}(2017)Springel, Pakmor, Pillepich, Weinberger, Nelson, Hernquist, Vogelsberger, Genel, Torrey, Marinacci, \& Naiman}]{TNG_4}
Springel, V., Pakmor, R., Pillepich, A., {et~al.} 2017, Monthly Notices of the Royal Astronomical Society, 475, 676–698, \dodoi{10.1093/mnras/stx3304}

\bibitem[{{Steidel}(1992)}]{Steidel_1992}
{Steidel}, C.~C. 1992, \pasp, 104, 843, \dodoi{10.1086/133065}

\bibitem[{Torrey {et~al.}(2019)Torrey, Vogelsberger, Marinacci, Pakmor, Springel, Nelson, Naiman, Pillepich, Genel, Weinberger, \& Hernquist}]{Torrey_2019}
Torrey, P., Vogelsberger, M., Marinacci, F., {et~al.} 2019, Monthly Notices of the Royal Astronomical Society, 484, 5587, \dodoi{10.1093/mnras/stz243}

\bibitem[{Truong {et~al.}(2020)Truong, Pillepich, Werner, Nelson, Lakhchaura, Weinberger, Springel, Vogelsberger, \& Hernquist}]{Truong_2020}
Truong, N., Pillepich, A., Werner, N., {et~al.} 2020, Monthly Notices of the Royal Astronomical Society, 494, 549, \dodoi{10.1093/mnras/staa685}

\bibitem[{Tumlinson {et~al.}(2017)Tumlinson, Peeples, \& Werk}]{Tumlinson_2017}
Tumlinson, J., Peeples, M.~S., \& Werk, J.~K. 2017, Annual Review of Astronomy and Astrophysics, 55, 389–432, \dodoi{10.1146/annurev-astro-091916-055240}

\bibitem[{van~de Voort {et~al.}(2019)van~de Voort, Springel, Mandelker, van den Bosch, \& Pakmor}]{van_de_Voort_2019}
van~de Voort, F., Springel, V., Mandelker, N., van den Bosch, F.~C., \& Pakmor, R. 2019, Monthly Notices of the Royal Astronomical Society: Letters, 482, L85, \dodoi{10.1093/mnrasl/sly190}

\bibitem[{{Verner} {et~al.}(1996){Verner}, {Ferland}, {Korista}, \& {Yakovlev}}]{Verner_1996}
{Verner}, D.~A., {Ferland}, G.~J., {Korista}, K.~T., \& {Yakovlev}, D.~G. 1996, \apj, 465, 487, \dodoi{10.1086/177435}

\bibitem[{Weinberger {et~al.}(2016)Weinberger, Springel, Hernquist, Pillepich, Marinacci, Pakmor, Nelson, Genel, Vogelsberger, Naiman, \& Torrey}]{TNG_6}
Weinberger, R., Springel, V., Hernquist, L., {et~al.} 2016, Monthly Notices of the Royal Astronomical Society, 465, 3291–3308, \dodoi{10.1093/mnras/stw2944}

\bibitem[{{Weiner} {et~al.}(2009){Weiner}, {Coil}, {Prochaska}, {Newman}, {Cooper}, {Bundy}, {Conselice}, {Dutton}, {Faber}, {Koo}, {Lotz}, {Rieke}, \& {Rubin}}]{Weiner_2009}
{Weiner}, B.~J., {Coil}, A.~L., {Prochaska}, J.~X., {et~al.} 2009, \apj, 692, 187, \dodoi{10.1088/0004-637X/692/1/187}

\bibitem[{Werk {et~al.}(2013)Werk, Prochaska, Thom, Tumlinson, Tripp, O'Meara, \& Peeples}]{Werk_2013}
Werk, J.~K., Prochaska, J.~X., Thom, C., {et~al.} 2013, The Astrophysical Journal Supplement Series, 204, 17, \dodoi{10.1088/0067-0049/204/2/17}

\bibitem[{{Weymann} {et~al.}(1979){Weymann}, {Williams}, {Peterson}, \& {Turnshek}}]{Weymann_1979}
{Weymann}, R.~J., {Williams}, R.~E., {Peterson}, B.~M., \& {Turnshek}, D.~A. 1979, \apj, 234, 33, \dodoi{10.1086/157470}

\bibitem[{{Wu} {et~al.}(2025){Wu}, {Cai}, {Lan}, {Zou}, {Anand}, {Dey}, {Li}, {Aguilar}, {Ahlen}, {Brooks}, {Claybaugh}, {de la Macorra}, {Doel}, {Ferraro}, {Forero-Romero}, {Gontcho A Gontcho}, {Honscheid}, {Juneau}, {Kehoe}, {Kisner}, {Lambert}, {Landriau}, {Le Guillou}, {Manera}, {Meisner}, {Miquel}, {Moustakas}, {Newman}, {Prada}, {Rossi}, {Sanchez}, {Schlegel}, {Schubnell}, {Siudek}, {Sprayberry}, {Tarl{\'e}}, {Weaver}, \& {Zou}}]{wu/etal:2025}
{Wu}, X., {Cai}, Z., {Lan}, T.-W., {et~al.} 2025, \apj, 983, 186, \dodoi{10.3847/1538-4357/adb28a}

\bibitem[{{York} {et~al.}(2000){York}, {Adelman}, {Anderson}, {Anderson}, {Annis}, {Bahcall}, {Bakken}, {Barkhouser}, {Bastian}, {Berman}, {Boroski}, {Bracker}, {Briegel}, {Briggs}, {Brinkmann}, {Brunner}, {Burles}, {Carey}, {Carr}, {Castander}, {Chen}, {Colestock}, {Connolly}, {Crocker}, {Csabai}, {Czarapata}, {Davis}, {Doi}, {Dombeck}, {Eisenstein}, {Ellman}, {Elms}, {Evans}, {Fan}, {Federwitz}, {Fiscelli}, {Friedman}, {Frieman}, {Fukugita}, {Gillespie}, {Gunn}, {Gurbani}, {de Haas}, {Haldeman}, {Harris}, {Hayes}, {Heckman}, {Hennessy}, {Hindsley}, {Holm}, {Holmgren}, {Huang}, {Hull}, {Husby}, {Ichikawa}, {Ichikawa}, {Ivezi{\'c}}, {Kent}, {Kim}, {Kinney}, {Klaene}, {Kleinman}, {Kleinman}, {Knapp}, {Korienek}, {Kron}, {Kunszt}, {Lamb}, {Lee}, {Leger}, {Limmongkol}, {Lindenmeyer}, {Long}, {Loomis}, {Loveday}, {Lucinio}, {Lupton}, {MacKinnon}, {Mannery}, {Mantsch}, {Margon}, {McGehee}, {McKay}, {Meiksin}, {Merelli}, {Monet}, {Munn}, {Narayanan}, {Nash}, {Neilsen}, {Neswold}, {Newberg}, {Nichol}, {Nicinski},
  {Nonino}, {Okada}, {Okamura}, {Ostriker}, {Owen}, {Pauls}, {Peoples}, {Peterson}, {Petravick}, {Pier}, {Pope}, {Pordes}, {Prosapio}, {Rechenmacher}, {Quinn}, {Richards}, {Richmond}, {Rivetta}, {Rockosi}, {Ruthmansdorfer}, {Sandford}, {Schlegel}, {Schneider}, {Sekiguchi}, {Sergey}, {Shimasaku}, {Siegmund}, {Smee}, {Smith}, {Snedden}, {Stone}, {Stoughton}, {Strauss}, {Stubbs}, {SubbaRao}, {Szalay}, {Szapudi}, {Szokoly}, {Thakar}, {Tremonti}, {Tucker}, {Uomoto}, {Vanden Berk}, {Vogeley}, {Waddell}, {Wang}, {Watanabe}, {Weinberg}, {Yanny}, {Yasuda}, \& {SDSS Collaboration}}]{York_2000}
{York}, D.~G., {Adelman}, J., {Anderson}, Jr., J.~E., {et~al.} 2000, \aj, 120, 1579, \dodoi{10.1086/301513}

\bibitem[{{Young} {et~al.}(1982){Young}, {Sargent}, \& {Boksenberg}}]{Young_1982}
{Young}, P., {Sargent}, W.~L.~W., \& {Boksenberg}, A. 1982, \apj, 252, 10, \dodoi{10.1086/159529}

\bibitem[{Zhu \& Ménard(2013)}]{Zhu_2013}
Zhu, G., \& Ménard, B. 2013, The Astrophysical Journal, 770, 130, \dodoi{10.1088/0004-637X/770/2/130}

\end{thebibliography}

\end{document}